\documentclass[
 amsmath, 
 aps, 
 pra,
 twocolumn,
]{revtex4-2}
\usepackage{graphicx}
\usepackage{dcolumn}
\usepackage{bm}
\usepackage{float}
\usepackage{hyperref}
\usepackage{siunitx}

\hypersetup{colorlinks = true, allcolors = blue}
\newcommand\numberthis{\addtocounter{equation}{1}\tag{\theequation}}

\begin{document}

\title{Ground-state cooling of multiple near-degenerate mechanical modes}

\author{Jin-Yu Liu$^{1}$}
\author{Wenjing Liu$^{1}$}
\author{Da Xu$^{1}$}
\author{Jia-Chen Shi$^{1}$}
\author{Qihuang Gong$^{1,2}$}
\author{Yun-Feng Xiao$^{1,2}$}
\email{yfxiao@pku.edu.cn}

\affiliation{$^{1}$State Key Laboratory for Mesoscopic Physics and Frontiers Science Center for Nano-optoelectronics, School of Physics, Peking University, 100871, Beijing, China\\
$^2$Collaborative Innovation Center of Extreme Optics, Shanxi University, Taiyuan 030006, China}

\date{\today}

\begin{abstract}
We propose a general and experimentally feasible approach to realize simultaneous ground-state cooling of arbitrary number of near-degenerate, or even fully degenerate mechanical modes, overcoming the limit imposed by the formation of mechanical dark modes.
Multiple optical modes are employed to provide different dissipation channels that prevent complete destructive interference of the cooling pathway, and thus eliminating the dark modes.
The cooling rate and limit are explicitly specified, in which the distinguishability of the optical modes to the mechanical modes is found to be critical for an efficient cooling process.
In a realistic multi-mode optomechanical system, ground-state cooling of all mechanical modes is demonstrated by sequentially introducing optical drives, proving the feasibility and scalability of the proposed scheme.
The work may provide new insights in preparing and manipulating multiple quantum states in macroscopic systems. 
\end{abstract}

\maketitle

Optomechanics \cite{Aspelmeyer2014}, exploring interactions between electromagnetic fields and mechanical vibrations, serves as an invaluable platform for studying macroscopic quantum phenomena such as macroscopic quantum coherence \cite{Ockeloen-Korppi2018,Riedinger2018,Ockeloen-Korppi2016,Massel2012,PhysRevLett.116.163602,Pepper2012,PhysRevA.89.014302} and classical-to-quantum transition \cite{PhysRevLett.88.120401,PhysRevLett.97.237201,POOT2012273}.
Application-wise, optomechanical sensors have demonstrated ultrahigh sensitivity in single particle sensing and precision measurements of displacements, forces, and accelerations \cite{Liu2020,Fogliano2021,Krause2012}. 
A premise of most of these applications is the ground-state cooling of the participating mechanical modes to suppress the thermal noise. 
However, thus far, though ground-state cooling has been investigated both theoretically \cite{PhysRevLett.80.688,PhysRevLett.99.093901,PhysRevLett.99.093902,PhysRevLett.110.153606,PhysRevA.91.033818} and experimentally \cite{Park2009,Rocheleau2010,PhysRevA.83.063835,Teufel2011,Chan2011,Verhagen2012,PhysRevLett.123.223602,Whittle1333} in a single mechanical mode, simultaneous cooling of multiple mechanical modes have not been demonstrated yet.
This seriously limits the applications of multi-mode optomechanical systems in quantum many-body simulation \cite{PhysRevLett.107.043603,PhysRevLett.111.073603,PhysRevLett.112.133604}, quantum information processing \cite{PhysRevLett.109.013603,PhysRevLett.107.133601,Okamoto2013,PhysRevLett.108.153603} and multiplexed sensing devices \cite{Truitt2007,Bargatin2012,Rabl2010,PhysRevLett.110.227202,PhysRevLett.117.017701}.

The major obstacle for multi-mode ground-state cooling is the formation of mechanical dark modes \cite{Genes2008,Sommer2019}.
As the mode density of states increases with the size of system, macroscopic resonators inevitably encounter multiple mechanical modes that are indistinguishable to the optical mode, i.e., their frequency differences become smaller than the optical linewidths. 
During the cooling processes, these modes hybridize and form dark modes that are decoupled from the optical field due to destructive interference \cite{Shkarin2014,Ockeloen-Korppi2019}, which prevents further cooling of the system.
So far, several methods have been theoretically proposed to break the dark modes, either lifting the degeneracy of the mechanical modes \cite{Lai2018,Zhang2019_dissipative} or inducing nonreciprocal energy flow \cite{Lai2020_cool,Habraken_2012,Kim2017,Xu2019}.
However, their realizations need either additional coupling structures \cite{Lai2018,Zhang2019_dissipative,Lai2020_cool,Habraken_2012,Kim2017} or sophisticated control over the frequency and phase of multiple optical pumps \cite{Xu2019} that are experimentally challenging.
\begin{figure}[t]
\includegraphics[width=8cm]{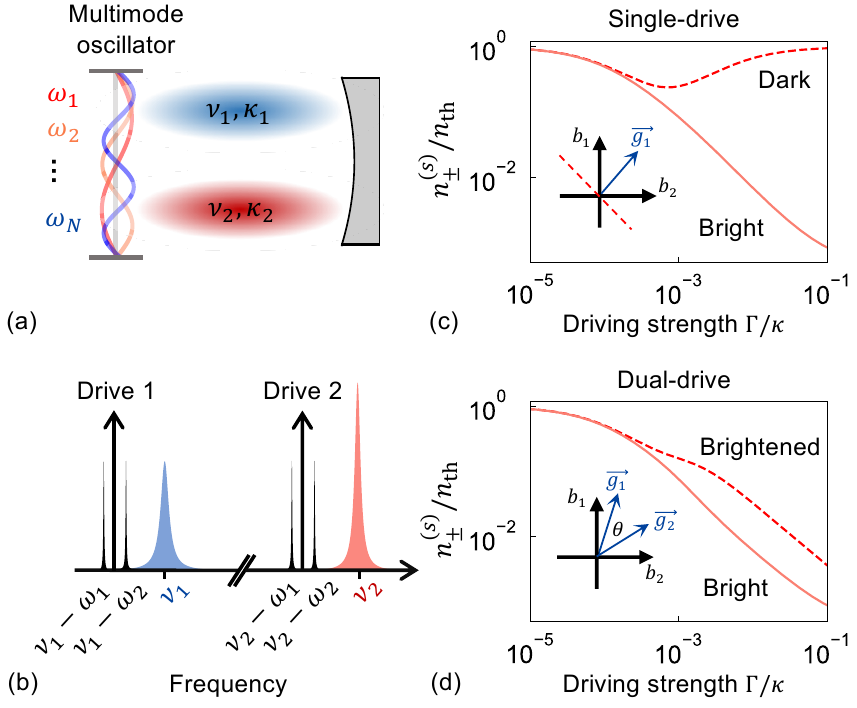}

\caption{(a) Schematic of mechanical resonator supporting multiple mechanical modes coupled to an optical cavity. 
(b) The optical drive configuration of cooling two mechanical modes via two optical modes.
(c)-(d) The normalized steady-state phonon number $n^{(s)}_\pm/n_\text{th}$ depending on $\Gamma/\kappa$ with single- and dual-drive, respectively. 
Here and hereafter, $\delta\omega_\text{mec}=0.001\kappa$, $\bar{\omega}_\text{mec}=20\kappa$.}
\label{fig1}
\end{figure}
Moreover, the complexity increases drastically with increasing number of mechanical modes as interaction engineering is required between each pair of the mechanical modes. 
In this Letter, we propose a new approach that is capable of simultaneous ground-state cooling of arbitrary number of  mechanical modes.
Multiple optical modes are implemented which effectively serve as different dissipation channels for phonons to break the destructive interference condition, and thereby preventing the formation of mechanical dark modes. 

As shown in Fig. \ref{fig1}(a), we consider that multiple mechanical modes with frequencies $\omega_j$ and linewidths $\gamma_j$ are coupled to multiple optical modes with frequencies $\nu_{k}$ and linewidths $\kappa_k$.
The optical modes are chosen to be well separated to prevent cross-mode interactions.
In the strong drive regime and in the rotation frame of the drive lasers, the linearized Hamiltonian of $N$ mechanical modes and $M$ optical modes reads (see Appendix \ref{A})
\begin{equation}\label{equ:H}
    H=\vec{a}^\dagger\boldsymbol{\Delta} \vec{a}+\vec{b}^\dagger\boldsymbol{\Omega} \vec{b}+(\vec{a}^\dagger \boldsymbol{G}\vec{b}+\vec{a}^T \boldsymbol{G}^*\vec{b}+\text{h.c.})
\end{equation}
where $\vec{a}=(a_1,\ a_2,\ ...a_M)^T$ is vector of linearized annihilation operators of the cavity modes, and $\vec{b}=(b_1,\ b_2,\ ...b_N)^T$ is vector of linearized annihilation operators of the mechanical modes. 
The diagonal matrix $\boldsymbol{\Omega}$ describes the frequencies of the $N$ mechanical modes, and
the diagonal matrix $\boldsymbol{\Delta}$ denotes the detunings of the $M$ drive lasers to their corresponding optical modes.
$\boldsymbol{G}$ is the linearized coupling matrix, with element $g_{kj}$ representing the coupling strength between the $k^{\text{th}}$ optical mode and $j^{\text{th}}$ mechanical mode. 
The optomechanical driving strength can be characterized by $\Gamma=\sum_{k=1}^M\Gamma_k$, with $\Gamma_k=\sum_{j=1}^N |g_{k,j}|^2/\kappa_k$ denoting the driving strength on the $k^\text{th}$ optical mode. 

We start the analysis with optomechanical cooling of two near-degenerate mechanical modes with frequencies $\omega_1$ and $\omega_2$, and linewidths $\gamma_1$ and $\gamma_2$, respectively.
To optimize the optomechanical cooling, the drive lasers are set to the resolved red sideband of the corresponding optical modes with detuning $\delta_k=(\omega_1+\omega_2)/2=\bar{\omega}_\text{mec}$ (see Appendix \ref{B}), as presented in Fig. \ref{fig1}(b).
To simplify the discussion, hereafter we assume $\kappa_1=\kappa_2=\kappa$, $\gamma_1=\gamma_2=\gamma=10^{-4}\kappa$, and identical driving strength $\Gamma_k$ of every optical mode.
The discussion of general systems parameters can be find in Supplemental Materials (see Appendix \ref{B}).
By performing adiabatic approximation to Eq. (\ref{equ:H}), the optomechanical interaction can be effectively understood as a $\Gamma$-dependent coupling between the mechanical modes that gives rise to two new mechanical eigenmodes $b_+$ and $b_-$.
As $\Gamma$ exceeds the mechanical frequency difference $\delta\omega_\text{mec}=|\omega_1-\omega_2|$, $b_\pm$ become the hybridization of $b_{1,2}$.
For a single optical drive, specifically, $b_+$ and $b_-$ can be written as $(g_{k,1}b_1+g_{k,2}b_2)/(g^2_{k,1}+g^2_{k,2})^{1/2}$ and $(-g_{k,2}b_1+g_{k,1}b_2)/(g^2_{k,1}+g^2_{k,2})^{1/2}$ when $\Gamma\gg\delta\omega_\text{mec}$.
In the Hilbert space spanned by $b_1$ and $b_2$, eigenmode $b_\pm$ correspond to the vectors along and perpendicular to the optomechanical coupling vector $\vec{g}_{k}=(g_{k,1},g_{k,2})$, respectively, as shown in the inset of Fig. \ref{fig1}(c).
Hence, $b_+$ is strongly coupled to the optical field and termed as the \textit{bright} mode, while $b_-$ is completely decoupled and termed as the \textit{dark} mode.

Figure \ref{fig1}(c) plots the steady-state phonon number of each mechanical eigenmode, $n^{(s)}_{\pm}=\langle b_{\pm}^\dagger b_{\pm} \rangle$, normalized to the thermal phonon number $n_\text{th}\approx(e^{\hbar\bar{\omega}_\text{mec}/kT}-1)^{-1}$.
{\textcolor{black}{Born-Markov approximation is applied in the calculation of steady-state phonon number here and hereafter, as no dominate mode exists in the thermal bath (see Appendix \ref{B}).}}
While both mechanical modes can be cooled at weak optical drives, from the onset of the mechanical strong coupling, the bright and dark mode starts to behave distinctly.
Although the bright mode is further cooled down, the dark mode is heated up due to its gradual decoupling from the optical modes \cite{Genes2008,Lai2020_cool}.
Particularly, $n^{(s)}_-$ approaches $n_\text{th}$ at $\Gamma\gg\delta\omega_\text{mec}$, indicating the complete suppression of the optomechanical cooling.
Such suppression acts as one of the major obstacles in the cooling of multi-mode mechanical oscillators, which has been widely observed in experiments \cite{Ockeloen-Korppi2019,Shkarin2014}. 
We propose this challenge can be resolved when a second optical mode is adopted for cooling, as seen in Fig. \ref{fig1}(d).
Upon the second coupling vector $\vec{g}_{k'}=(g_{k',1},\ g_{k',2})$ is introduced, as long as the two coupling vectors $\vec{g}_{k}$ and $\vec{g}_{k'}$ are not collinear ($\theta\neq 0$), no mechanical mode can be decoupled from both optical modes.
Indeed, in this case, both mechanical modes can be efficiently cooled either before or after the mechanical strong coupling.

\begin{figure}[bt]
\includegraphics[width=8.5cm]{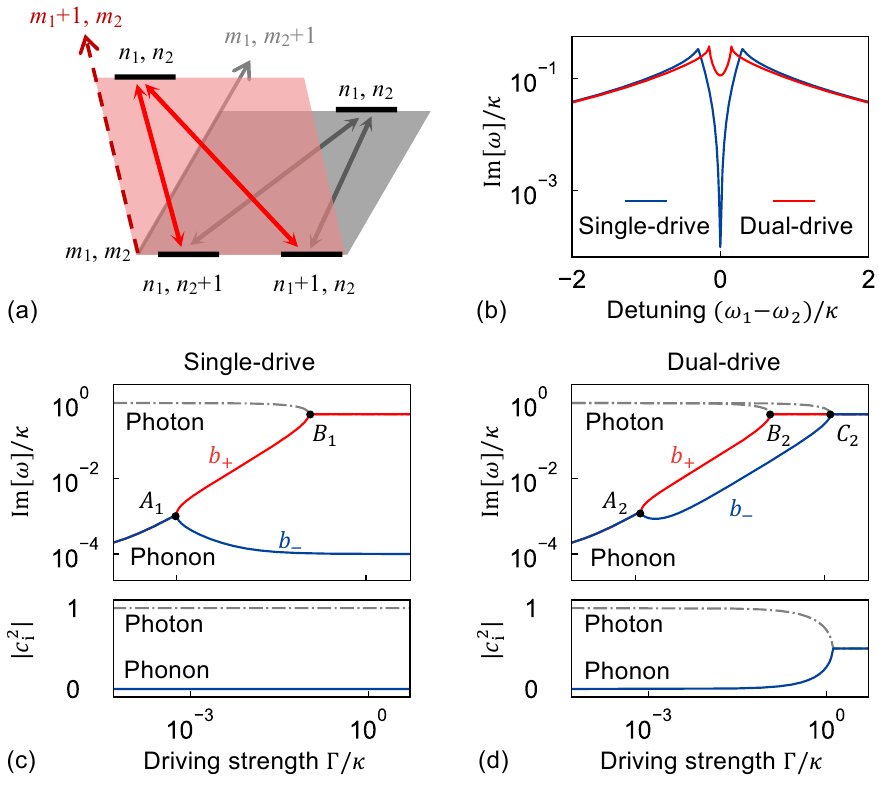}
\caption{(a) Schematic of the energy transfer pathways of two mechanical modes coupled with two optical drives. 
(b) Dissipation spectra of the mechanical modes under the single- and dual-drive  conditions, both with $\Gamma=0.1\kappa$. 
(c)-(d) The imaginary part of the eigenvalues $\text{Im}[\omega]/\kappa$ and the composition of the $b_-$ mode, as a function of $\Gamma/\kappa$ with single- (c) and double- (d) drive.
The gray dash-dotted line, the solid red line, and the solid blue lines denote the optical, $b_+$, and $b_-$ modes. Here $\cos\theta=0.8$.}
\label{fig2}
\end{figure}

From the energy transfer aspect, different optical modes serve as different dissipation channels for phonons in optomechanical cooling. With a single optical drive, near-degenerate phonon modes decaying through the same channel interfere destructively with each other, analogous to the electromagnetic induced transparency (EIT), as shown in Fig. \ref{fig2}(a) \cite{PhysRevLett.66.2593}. 
Within the transparency window, the mechanical mode is decoupled from the optical field with its damping rate reduced to the intrinsic linewidth $\gamma$ (Fig. \ref{fig2}(b)).
When multiple optical pathways present, phonon dissipation forbidden in one pathway can decay through another, which effectively removes the EIT window and brighten up the dark mode.

The evolution of the mechanical dark mode can be quantitatively investigated by the eigenvalues and eigenvectors of the system, as shown in Figs. \ref{fig2}(c) and (d). 
At $\Gamma\approx\delta\omega_\text{mec}$, exceptional points present at $A_1$ and $A_2$ in the single- and dual- drive schemes, respectively, indicating the formation of the bright and dark mechanical modes.
In both cases, the bright modes $b_+$ (red curves) exhibit rapid dissipation rate increases, quickly reach their classical cooling limits at the second exceptional points of the strong optomechanical coupling, denoted by $B_1$ and $B_2$, respectively.
On the other hand, right after the exceptional points $A_\text{1}$ and $A_\text{2}$, the dissipation rate of both $b_-$ modes (blue curves) decrease with the driving strengths.
In the single-drive scheme, it decreases monotonically to the intrinsic dissipation of the mechanical mode $\gamma$ hence that the cooling is completely suppressed.
Oppositely, in the proposed dual-drive scheme, the $b_-$ mode is brightened up and eventually reaches an emerging exceptional point $C_2$. 
Meanwhile, by examining the eigenvectors of the mechanical modes, one can see that in the single-drive scheme, the mechanical dark mode becomes purely phononic with strong optical drive, while in the dual-drive scheme, it is significantly hybridized with the photonic modes. 
The brightened $b_-$ mode exhibits its classical cooling limit when reaching point $C_2$, which is characterized by $n^{(s)}_-=\gamma n_\text{th}/(\gamma+\kappa)$.
This value recovers the cooling limit of the single-mechanical mode system (see Appendix \ref{D}), demonstrating the elimination of the dark mode effect.
It should also be noted that the quantum cooling limit of the multi-mode system can be estimated as $\kappa^2/(16\bar{\omega}_\text{mec}^2)$ (see Appendix \ref{E}), and at the condition under investigation, it is 2 orders of magnitude smaller than the classical limit.

While the restriction on the cooling limit can be lifted by any drive configurations with noncollinear coupling vectors, a further optimization of the system parameters that can minimize the required driving strength is important for experimental realizations.
Quantitatively, the steady-state total phonon number $n^\text{(s)}_\text{tot}=n^{(s)}_++n^{(s)}_-$ at a given driving strength $\Gamma$ is calculated (see Appendix \ref{B}), in the drive range where the bright and dark modes are formed but the system still remains in the weak coupling regime,
\begin{align*}\label{align*:ntot}
        n^\text{(s)}_\text{tot} =& \frac{2\gamma n_\text{th}(\gamma+2\Gamma)}{\gamma^2+4\gamma\Gamma+4\Gamma^2\sin^2\theta}  \numberthis .
\end{align*}

It can be seen that the angle $\theta$ is the key parameter for achieving efficient cooling, which describes the distinguishability of the optical modes to the mechanical modes. 
At $\Gamma\gg\gamma$, when $\theta=0$, $n^\text{(s)}_\text{tot}\approx n_\text{th}$, the system is equivalent to be driven by a single optical pump and the cooling is suppressed. 
When $\theta\neq0$, $n^\text{(s)}_\text{tot}\approx n_\text{th} \gamma/({\Gamma\sin^2\theta})$, indicating that the cooling is more efficient when the chosen optical modes exhibit more distinct coupling strengths to the mechanical modes.
Especially, when $\theta=\pi/2$, $n_\text{tot}$ reaches the minimum at a given $\Gamma$.
In this case, each drive interacts exclusively with the $b_+$ or $b_-$ mode, and the system can be reduced to a single-mechanical mode resonator. 

\begin{figure}[tb]
\includegraphics[width=8.5cm]{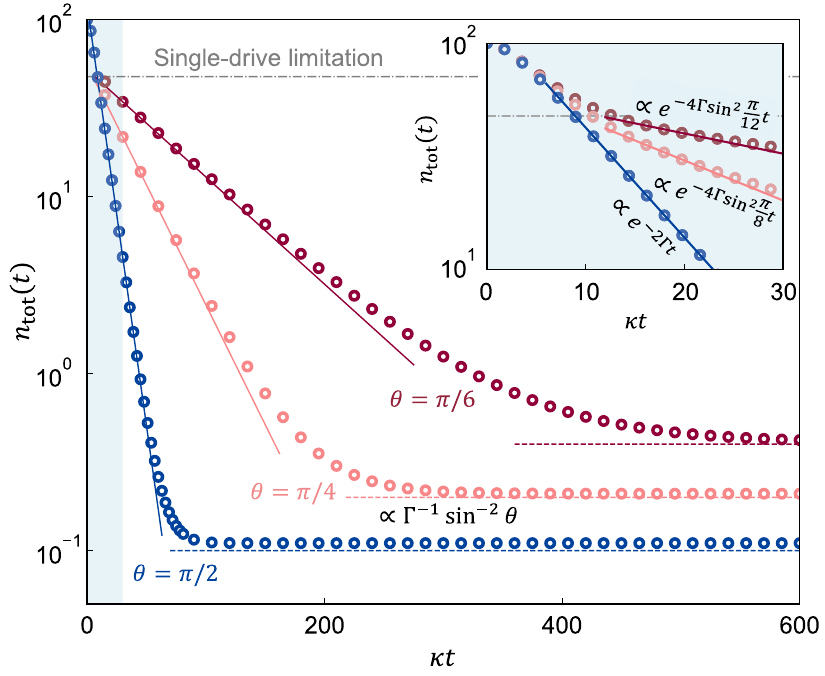}
\caption{Temporal evolution of the total phonon number $n_\text{tot}(t)$ during the cooling process.
Hollow circles denote the numerically calculated phonon number; solid- and dashed-lines represent the analytically calculated phonon decay rate and the steady-state phonon number, respectively; dash-dotted line denotes the steady-state phonon number cooled by a single-optical drive. 
Inset: the zoomed-in plot of $\kappa t=[0, 30]$, represented by the blue shaded region in the main figure.
Here $\Gamma=0.05\kappa$ and $n_\text{tot}(0)=100$.}
\label{fig3}
\end{figure}

The cooling speed is another important figure of merit for optomechanical cooling processes. 
Here, the temporal evolution of the total phonon number $n_\text{tot}(t)$ is calculated numerically with the $4^\text{th}$-order Runge-Kutta method, as shown in Fig. \ref{fig3}.
The optical drives are turned on at $\kappa t=0$ and kept constant at $\Gamma=0.05\kappa$.
For $0<\theta<\pi/2$, $n_\text{tot}(t)$ is characterized by a double-exponential decay, with the fast and slow processes, $\exp[-4\Gamma\cos^2(\theta/2)t]$ and $\exp[-4\Gamma\sin^2(\theta/2)t]$ corresponding to the $b_+$ and $b_-$ modes, respectively.
At small $\kappa t$, the system exhibits a rapid phonon dissipation of both mechanical modes, characterized by a $\theta$-independent decay rate of $\Gamma$, till cooling limitation of the single-drive configuration (gray dotted line), as shown in the inset of Fig. \ref{fig3}.
Below the single-drive limitation line, most phonons in the $b_+$ mode have already been dissipated from the system due to the larger decay rate, and thus the process is dominated by the $b_-$ mode. 
In this regime, a large $\theta$ results in a significantly accelerated phonon decay rate.
In particular, for the case $\theta=\pi/2$, the system reduces to the single-mechanical mode cooling case with a mono-exponential decay rate of $\Gamma$, until reaching the steady state.
Meanwhile, the steady-state phonon number $n_\text{tot}$ also decreases while increasing $\theta$, and the numerical results match well with the analytical solution described by Eq. (\ref{align*:ntot}).

Finally, we present here that our method can be straightforwardly generalized to arbitrary number of mechanical modes.
In the $N$ dimensional Hilbert space of mechanical modes, a mode is dark if it is orthogonal to all the coupling vectors $\vec{g}_k=(g_{k,1},\ g_{k,2},\ ...g_{k,N})$. 
Hence, a dark subspace can be defined as the orthogonal complement of the span of all coupling vectors.
When $M$ optical modes with linearly independent coupling vectors are introduced, the dimension of dark subspace is reduced to $N-M$. 
Given that $M\geq N$, all dark modes are eliminated and the ground-state cooling for $N$ degenerate mechanical modes can be achieved.
\begin{figure}[tb]
\includegraphics[width=8cm]{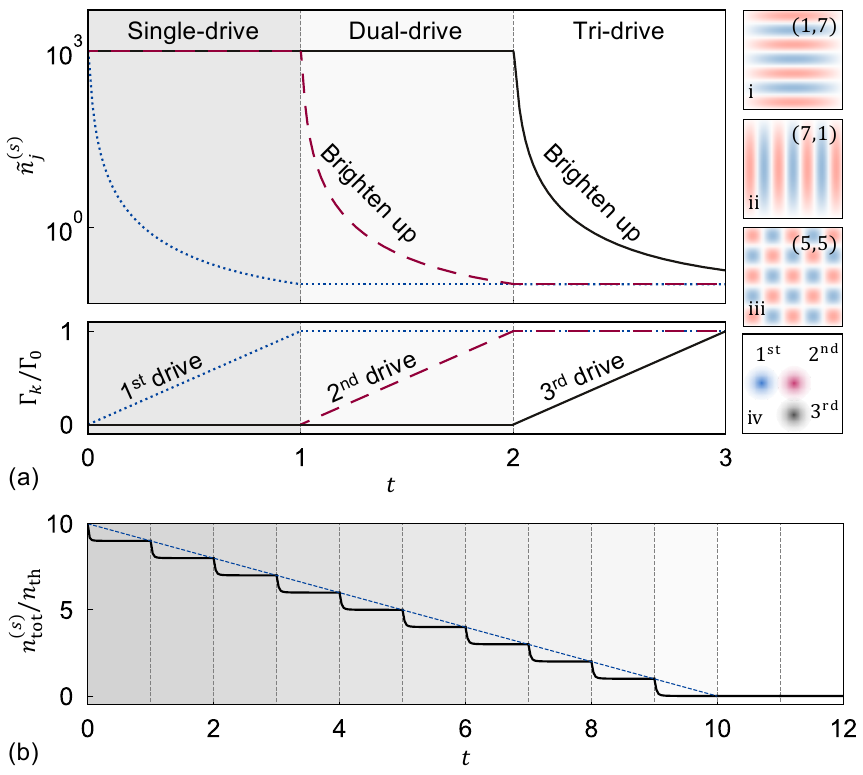}
\caption{(a) Evolution of the steady phonon number $\tilde{n}_j^{(s)}$ of each mechanical eigenmode (upper panel), as three optical drives are quasi-statically introduced (lower panel). 
The mechanical modes are fully degenerate with $\omega_j/2\pi=\SI{1.178}{\mega\hertz}$, $\gamma_j/2\pi=\SI{39.0}{\milli\hertz}$, and their mode profiles are presented in i-iii; Optical modes have the linewidths of $\kappa_k/2\pi=\SI{0.967}{\mega\hertz}$ and $\Gamma_0=\SI{980}{Hz}$. The spatial profiles are shown in iv. (b) Evolution of $n^{(s)}_\text{tot}$ of a $N=10$ system as 
{\textcolor{black}{$12$}} optical drives introduced sequentially.
In each regime divided by the vertical dashed lines, one optical drive is turned on and increased quasi-statically from $0$ to $\Gamma_0=0.1\kappa$.
}
\label{fig4}
\end{figure}
As an example, a silicon nitride membrane with clamped boundaries inserted into a Fabry-P$\acute{\text{e}}$rot optical cavity is considered, with all the optomechanical parameters corresponding to realistic experimental systems \cite{Purdy_2012,Jayich_2008,PhysRevLett.108.083603,PhysRevLett.116.147202}.
Cooling of the three-fold degenerate mechanical drum modes, $(1,7)$, $(7,1)$, and $(5,5)$ is demonstrated, with three spatially distinct optical modes employed to allow large $\theta$, as shown in Fig. \ref{fig4}(a) i-iv.
{\textcolor{black}{Such drive scheme can be realized by focusing the laser drive onto different positions of the membrane and adjusting the membrane position in the cavity \cite{Purdy_2012}}}.
The three optical drives are turned on sequentially in the quasi-static limit to examine the phonon number evolution of the three mechanical modes.
When the first drive is turned on and its strength $\Gamma_1$ is increased from 0 to $\Gamma_0$, the three mechanical modes hybridized to form one bright mode ($\tilde{b}_1$, blue curve) and two dark modes ($\tilde{b}_2$ and $\tilde{b}_3$, black and magenta curves, respectively).
Once turning on the second drive, one of the dark modes $\tilde{b}_2$ is brightened up, leaving only one dark mode in the system.
All three modes are effectively turned bright and cooled down to $\tilde{n}^{(s)}_j<1$ when the third drive is on, achieving simultaneous ground-state cooling of all three mechanical modes under investigation.
{\textcolor{black}{We note that the cooling is independent of the drive sequence as the system is linearized with a unique steady state, e.g., the steady state phonon number remains the same when all three drives are induced at the same time (see Appendix \ref{C}).}}

For a more general demonstration, the analysis is further pushed to the cooling process of more mechanical modes, for example $N=10$. With {\textcolor{black}{$12$}} drives quasi-statically introduced and linearly enhanced in sequence, the result is shown in Fig. \ref{fig4}(b).
{\textcolor{black}{When $M<10$}}, in the presence of a new optical drive, $n^{(s)}_\text{tot}$ undergoes a pronounced decrease and reaches a cooling limit at $n^{(s)}_\text{tot}\approx (10-M)n_\text{th}$, {\textcolor{black}{if neglecting the phonon occupancy of the bright modes,}} as represented by the blue dashed line. 
Such step-like cooling curve indicates the successive elimination of the dark modes by each optical drive, as predicted by the theory.
{\textcolor{black}{When $M\geq10$, all dark modes have been eliminated, and further introduction of additional optical modes no longer leads to significant cooling other than the increase of the total drive strength.}}

As for the cooling limit, when $M$ drives with $M=N$ are employed and all the dark modes are eliminated, $n^{(s)}_\text{tot}$ in weak coupling regime can be asymptotically described by 
(see Appendix \ref{B}) 
\begin{equation}
n^{(s)}_\text{tot}=\frac{1}{4}\gamma\kappa n_\text{th}\parallel \!\boldsymbol{G}^{-1}\!\!\parallel_2,
\end{equation}
where {\textcolor{black}{$\parallel\!\! \boldsymbol{G^{-1}}\!\parallel_2 =\sum_{k,j=1}^N|(g^{-1})_{k,j}|^2$}}. 
Here, $\boldsymbol{G}^{-1}$ exists if and only if all the coupling vectors are linearly independent.  
For each $\vec{g}_k$, the rest coupling vectors span an $N-1$ dimensional hyper surface in the $N$ dimensional Hilbert space. 
For linearly independent coupling vectors, the cross angle $\theta_k$ between $\vec{g}_k$ and this hyper surface is nonzero, and the cooling limit can be rewritten as (see Appendix \ref{B})
\begin{equation}
n^{(s)}_\text{tot}=\frac{\gamma n_\text{th}}{4}\sum_{k=1}^N \frac{1}{\Gamma_k\sin^2\theta_k}\quad (\theta_k\neq 0).
\end{equation}
Hence a larger $\sin^2\theta_k$ can result in a better cooling performance. Also, if all $\Gamma_k$ are kept constant, the best cooling is achieved when all $\theta_k=\pi/2$ for $k=1,\ 2,\ ...N$. In this case, each optical mode solely couples to one mechanical mode.
This result provides a quantitative guidance for selecting the optical modes that are best suitable for cooling a multi-mechanical mode system.

In conclusion, we have proposed a general scheme to realize the ground-state cooling of near-degenerate or even degenerate mechanical modes. 
Different optical modes provide different dissipation channels that can effectively eliminate the mechanical dark modes that obstruct the cooling process.
The distinguishability of the optical modes to the mechanical modes is found to be an essential factor that allows efficient optomechanical cooling.
This approach not only provides an experimental feasible method that may help to solve one of the critical challenges in fundamental and applied studies on macroscopic optomechanics, but could also inspire dark mode manipulation and suppression in analogous systems such as cold atom ensembles.

J.-Y. Liu, W. Liu, and D. Xu contributed equally. This project is supported by the National Key R\&D Program of China (Grant No. 2018YFA0704404) and the National Natural Science Foundation of China (Grants Nos. 11825402, 11654003, 61435001, and 62035017).

\appendix{Supplemental Material}

\section{Derivation of the linearized Hamiltonian of a multi-mode optomechanical system}\label{A} 
The Hamiltonian of an optically driven multi-mode optomechanical system can be generally written as
\begin{equation}
    H=H_\text{free}+H_\text{int}+H_\text{drive}.
\end{equation}
Here, $H_\text{free}$ is the free Hamiltonian of the optical and mechanical modes with
\begin{equation}
    H_\text{free}=\sum_{k=1}^M\hbar\nu_ka_k^\dagger a_k+\sum_{j=1}^N\hbar\omega_{j}b_j^\dagger b_j,
\end{equation}
where $a_k (a_k^\dagger)$ is the annihilation (creation) operator of the $k^\text{th}$ optical mode with frequency $\nu_k$ and linewidth $\kappa_k$; $b_j (b_j^\dagger)$ is the annihilation (creation) operator of the $j^\text{th}$ mechanical mode with frequency $\omega_j$ and linewidth $\gamma_j$. 
These operators obey the bosonic commutation relations
\begin{align}
[a_{k},a_{k'}^\dagger ]=\delta_{kk'},\ 
[ b_j ,b^\dagger_{j'} ]=\delta_{jj'},\  
[ a_k,b_j^{(\dagger)}]=0.
\end{align}
$H_\text{int}$ represents the interaction Hamiltonian between the optical and mechanical modes \cite{PhysRevA.51.2537}
\begin{equation}
    H_\text{int}=\hbar\sum_{k=1}^M\sum_{j=1}^N g^S_{k,j}a_k^\dagger a_k(b_j^\dagger+ b_j),
\end{equation}
where $g^S_{k,j}$ is the single photon coupling strength between the $k^\text{th}$ optical mode and the $j^\text{th}$ mechanical mode.

\hfill

\noindent
$H_\text{drive}$ describes the laser drive on the optical modes
\begin{equation}
    H_\text{drive}=\sum_{k=1}^M Q_{k}e^{-i\omega^\text{d}_{k} t} a_k^\dagger+ \text{h.c. },
\end{equation}
where $Q_k$ is the driving amplitude and $\omega^\text{d}_k$ is the driving frequency for the $k^\text{th}$ optical mode. 

In the rotating frames of the drive lasers
$S[t]=\exp{[-it\sum \omega^\text{d}_{k} a_k^\dagger a_k]}$,
the optical operators are transformed to $
    S^\dagger[t] a_kS[t]=a_ke^{-i\omega^\text{d}_{k} t}
    $. 
The Hamiltonian is transformed to $H'=S^\dagger [t] H S[t]-i\hbar\partial_t S[t]$ and written as
\begin{align}
    &H_\text{free}'=\sum_{k=1}^M\hbar\delta'_{k}a_k^\dagger a_k+\sum_{j=1}^N\hbar\omega_{j}b_j^\dagger b_j,\\
    &H_\text{int}'=H_\text{int},\\
    &H_\text{drive}'=\sum_{k=1}^M Q_{k}a_k^\dagger+ \text{h.c. },
\end{align}
where the original drive detunning $\delta'_k=\nu_k-\omega^\text{d}_{k}$.
{\textcolor{black}{With Born-Markov approximation,}} the Langevin equations of the system can therefore be written as
\begin{align}
    \frac{\text{d}a_k}{\text{d}t}=(&-i\delta'_k-\frac{\kappa_k}{2})a_k-i\sum_{j=1}^N g^S_{k,j}a_k(b_j^\dagger+b_j)\nonumber\\&-iQ_k-\sqrt\kappa_k a^\text{in}_k(t),\\
    \frac{\text{d}b_j}{\text{d}t}=(&-i\omega_j-\frac{\gamma_j}{2})b_j-i\sum_{k=1}^M g^S_{k,j}a_k^\dagger a_k-\sqrt{\gamma_j}b^\text{in}_j(t),
\end{align}
where $a^\text{in}_k,b^\text{in}_j$ are the input operators of the optical and mechanical modes, which obey
\begin{align}
    \langle {a^\text{in}_k}^\dagger(t_k) a^\text{in}_{k'}(t_{k'})\rangle&=\delta_{kk'}\delta(t_k-t_{k'})n_\text{th}(\nu_k),\\
    \langle {b^\text{in}_j}^\dagger(t_j)
    b^\text{in}_{j'}(t_{j'})\rangle&=\delta_{jj'}\delta(t_j-t_{j'})n_\text{th}(\omega_j).
\end{align}
Here the thermal noise $n_\text{th}(\omega)=1/(e^{\hbar\omega/kT}-1)$ with $T$ being the environment temperature. 
{\textcolor{black}{The thermal bath is composed of numerous free-space electromagnetic modes, which all couple weakly to the system. 
As no dominant mode exists in the thermal bath, the back-action on thermal bath can be ignored with the Markov approximation applicable. 
}}As the high optical frequency condition $\hbar\nu_k\gg kT$ applies for common experimental conditions, thermal noise $n_\text{th}(\nu_k)\approx0$ for the optical modes.
The mechanical modes are near degenerate with $\omega_{j}\approx\bar{\omega}_\text{mec}=(\sum_{j=1}^N\omega_j)/N$, thus all $n_\text{th}(\omega_j)\approx n_\text{th}(\bar{\omega}_\text{mec})$, labeled as $n_\text{th}$ hereafter. 
The remaining independent quadratic expressions of ${a^\text{in}_k}^{(\dagger)},{b^\text{in}_j}^{(\dagger)}$ have zero value expectations.

The operators $o\in\{a_k^{(\dagger)},\ b_j^{(\dagger)}\}$ can be divided into their expectations and fluctuations, $o=\langle o\rangle+\delta o$. Defining $\alpha_k=\langle a_k\rangle$ and $\beta_j=\langle b_j\rangle$, the Langevin equations are split into equations of the expectations

\begin{align}
    \frac{\text{d}\alpha_k}{\text{d}t}&=(-i\delta'_k-\frac{\kappa_k}{2})\alpha_k-i\sum_{j=1}^N g^S_{k,j}\alpha_k(\beta_j^*+\beta_j)-iQ_k,\\
    \frac{\text{d}\beta_j}{\text{d}t}&=(-i\omega_j-\frac{\gamma_j}{2})\beta_j-i\sum_{k=1}^M g^S_{k,j}\alpha_k^* \alpha_k,
    \end{align}
and equations of the fluctuations
    \begin{align}
        &\frac{\text{d}\delta a_k}{\text{d}t}=(-i\delta_k-\frac{\kappa_k}{2})\delta a_k\nonumber\\
    &-i\sum_{j=1}^N g_{k,j}(\delta b_j^\dagger+\delta b_j)-\sqrt\kappa_j a^\text{in}_j(t),\\
    &\frac{\text{d}\delta b_j}{\text{d}t}=(-i\omega_j-\frac{\gamma_j}{2})\delta b_j\nonumber\\
    &-i\sum_{k=1}^M(g_{k,j}\delta a_k^\dagger+g_{k,j}^*\delta a_k)-\sqrt{\gamma_j}b^\text{in}_j(t)\label{langevin_mec},
\end{align}

respectively, where the corrected drive detuning and the linear coupling strength are defined as $\delta_k=\delta_k'+\sum_{j=1}^N g_{k,j}(\beta_j^*+\beta_j)$ and $g_{k,j}=g^S_{k,j}\alpha_k $, respectively, for the $k^\text{th}$ optical mode and the $j^\text{th}$ mechanical mode.
The steady-state expectations of the optical and mechanical modes at $\text{d}\langle  o^{(\dagger)} \rangle/\text{d}t=0$ are
\begin{align}
    \alpha_k&=\frac{-iQ_k}{i\delta_k+\kappa_k/2},\\
    \beta_j&=\frac{i\sum_{k=1}^M g^S_{k,j}|\alpha_k|^2}{i\omega_j+\gamma_j/2}.
\end{align}
The above Langevin equations of fluctuations are equivalent to the linearized Hamiltonian
\begin{align}
    &H_L\!\!=\!\!\sum_{j=1}^N\omega_j\delta b_j^\dagger \delta b_j+\!\!\sum_{k=1}^M\!\delta_k\delta a_k^\dagger \delta a_k\\\nonumber
    &+\sum_{j=1}^N\sum_{k=1}^M\!\!\left[g_{k,j}\delta a_k^\dagger (\delta b_j+\delta b_j^\dagger)\!+\! \text{h.c.}\!\right].
\end{align} 
In the main text and the following sections of the Supplemental material, all the symbols $(\delta a,\ \delta b)$ are relabeled as $(a,\ b)$ for simplicity.

\section{Calculation of the steady-state phonon number}\label{B} 
The steady-state phonon number is calculated via the Lyapunov equation derived from the quantum master equation \cite{Praxmeyer_2019}. 
The equations read
\begin{widetext}
\begin{align}
    \label{equ:1}\frac{\text{d}}{\text{d}t}\langle b_{j'}^\dagger b_j\rangle\!=&(i\omega_{j'}\!-i\omega_{j}\!-\frac{\gamma_{j'}+\gamma_j}{2})\langle b_{j'}^\dagger b_j\rangle+\gamma_{j}\delta_{jj'}n_\text{th}
    \!-i\sum_{k=1}^M\langle b_{j'}^\dagger(g_{k,j}^* a_k+g_{k,j} a_k^\dagger)\rangle+i\sum_{k=1}^M\langle(g_{k,j'}^* a_k+g_{k,j'} a_k^\dagger) b_j\rangle,\\
    \label{equ:1'}\frac{\text{d}}{\text{d}t}\langle b_{j'} b_j\rangle\!=&(-i\omega_j-i\omega_{j'}-\frac{\gamma_{j}+\gamma_{j'}}{2})\langle b_{j'} b_j\rangle
    -i\sum_{k=1}^M\langle b_{j'}(g_{k,j}^* a_k+g_{k,j} a_k^\dagger)\rangle
    -i\sum_{k=1}^M\langle(g_{k,j'}^* a_k+g_{k,j'} a_k^\dagger) b_j\rangle,\\
    \label{equ:2}
    \frac{\text{d}}{\text{d}t}\langle a_{k'}^\dagger a_{k}\rangle\!=&(i\delta_{k'}-i\delta_{k}-\frac{\kappa_{k}+\kappa_{k'}}{2})\langle a_{k'}^\dagger a_{k}\rangle
    -i\sum_{j=1}^N g_{k,j}\langle a_{k'}^\dagger( b_j+ b_j^\dagger)\rangle+i\sum_{j=1}^N g_{k',j}^*\langle( b_j+ b_j^\dagger) a_{k}\rangle,\\
    \label{equ:2'}\frac{\text{d}}{\text{d}t}\langle a_{k'} a_{k}\rangle\!=&(-i\delta_{k'}-i\delta_{k}-\frac{\kappa_{k'}+\kappa_k}{2})\langle a_{k'} a_{k}\rangle
    -i\sum_{j=1}^N g_{k,j}\langle a_{k'}( b_j+ b_j^\dagger)\rangle-i\sum_{j=1}^N g_{k',j}\langle( b_j+ b_j^\dagger) a_{k}\rangle,\\
    \label{equ:3}
    \frac{\text{d}}{\text{d}t}\langle a_k^\dagger b_j\rangle\!=&(i\delta_k-i\omega_{j}-\frac{\gamma_{j}+\kappa_k}{2})\langle a_k^\dagger b_j\rangle
    -i\sum_{k'=1}^M\langle a_k^\dagger(g_{k',j}^* a_{k'}+g_{k',j} a_{k'}^\dagger)\rangle+i\sum_{j'=1}^N g_{k,j'}^*\langle( b_{j'}+ b_{j'}^\dagger) b_{j}\rangle,\\
    \label{equ:3'}\frac{\text{d}}{\text{d}t}\langle a_k b_j\rangle\!=&(-i\delta_k-i\omega_{j}-\frac{\gamma_{j}+\kappa_k}{2})\langle a_k b_j\rangle-ig_{k,j}
    -i\sum_{k'=1}^M\langle (g_{k',j}^* a_{k'}+g_{k',j} a_{k'}^\dagger)a_k \rangle-i\sum_{j'=1}^N g_{k,j'}\langle( b_{j'}+ b_{j'}^\dagger) b_j\rangle.
\end{align}
\end{widetext}
Under the steady-state condition $\text{d}\langle o^{(\dagger)}  o'\rangle/\text{d}t=0$, the system can be solved by a 36-dimensional linear equation. 
The steady-state phonon number of a mechanical mode $\tilde{b}_j=\sum_{j'=1}^N e_{jj'}b_{j'}$ is written as
\begin{align}\label{align:ns}
    \tilde{n}_j^{(s)}&=\langle\tilde{b}_j^\dagger\tilde{b}_j\rangle=\sum_{j'=1}^N\sum_{j''=1}^N e_{jj'}^*e_{jj'}\langle b_{j'}^\dagger b_{j''}\rangle
\end{align}

\begin{figure}[h]
\includegraphics[width=8.5cm]{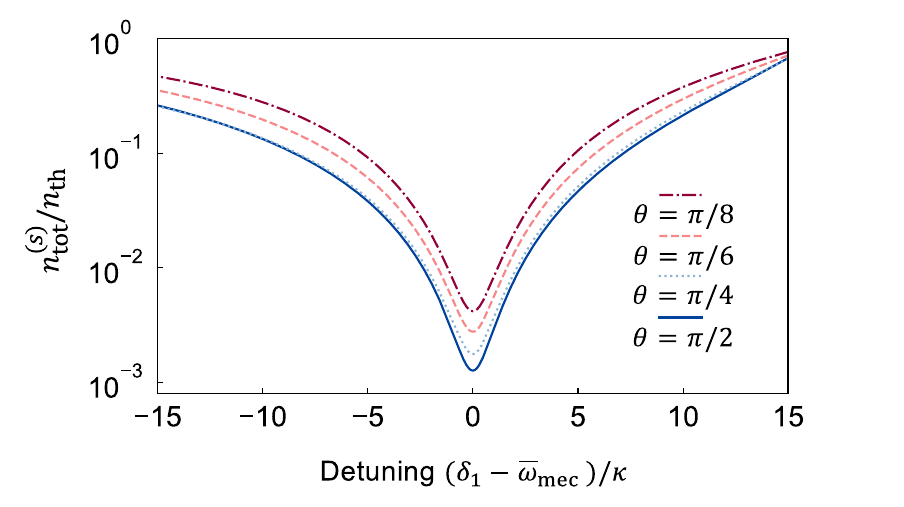}
\caption{Total steady-state phonon number $n^{(s)}_\text{tot}$ as a function of detuning $\delta_1$ between the $1^\text{st}$ optical mode and its drive. Here $N=M=2$, $\Gamma_1=\Gamma_2=0.1$, $\delta_2=\bar{\omega}_\text{mec}$. Rest of the parameters are the same as Fig. \ref{fig1} of the main text.} 
\label{figSI1}
\end{figure}

To optimize the cooling process of a multi-mode resonator, the laser drive detuning is scanned across the red sideband of the optical mode while calculating the steady-state total phonon number $n^{(s)}_\text{tot}=\sum_{j=1}^N\langle b_j^\dagger b_j\rangle$, as shown in Fig. \ref{figSI1}, which indicates that the best cooling performance is reached at $\delta_k=\bar{\omega}_\text{mec}$.

\noindent 

This optimized driving condition and Eq. (\ref{align:ns}) give rise to the numerically calculated steady-state phonon numbers presented in Figs. \ref{fig1} and \ref{fig1} in the main text.

Analytically, further applying the rotating wave and the adiabatic approximation in the weak coupling regime $(g_{k,j}\ll\kappa_{k'})$, terms $\langle o o'\rangle^{(*)}$ can be ignored, and the Fourier transformed Langevin equations of $a_k$ writes
\begin{align}\label{eqn:a(omega)}
     a_k(\omega)&=\frac{-i\sum_{j=1}^N g_{k,j} b_j(\omega)-\sqrt\kappa_k a^\text{in}_k(\omega)}{i\delta_k-i\omega+\frac{\kappa_k}{2}}.
\end{align}
When returning to the time domain, the difference between $b_j(\omega)$ and $b_j(\omega_j)\delta(\omega-\omega_j)$ can be ignored for $g_{k,j}\ll \kappa_{k'}$, thsu $a_k$ is expressed as
\begin{align}
     a_k&=\sum_{j=1}^N\frac{-i g_{k,j} b_j}{i\delta_k-i\omega_{j'}+\frac{\kappa_k}{2}}.\label{no_thermal_opt}
\end{align}
Assuming that the mechanical modes are optically indistinguishable $(|\omega_j-\omega_{j'}|\ll\kappa_k)$, and that all the drives are located at the red sideband with $\delta_k=\bar{\omega}_\text{mec}\approx \omega_{j}$, Eq. (\ref{no_thermal_opt}) can be simplified as
\begin{align}
     a_k&=\sum_{j=1}^N\frac{-2ig_{k,j} b_j}{\kappa_k}.\label{adrwa_opt}
\end{align}
After substituting Eq. (\ref{adrwa_opt}) into Eq. (\ref{langevin_mec}), the later becomes
\begin{align}
    \frac{\text{d} b_j}{\text{d}t}=(&-i\omega_{j}-\frac{\gamma_{j}}{2}) b_j
    -2\sum_{j'=1}^N\sum_{k=1}^M\frac{g_{k,j}^*g_{k,j'} b_{j'}}{\kappa_k}-\sqrt{\gamma_{j}}b^\text{in}_j(t).\label{rw_temporal}
\end{align}
The second term on the right-hand side of Eq. (\ref{rw_temporal}) represents the optically induced damping, of which the Hermitian operator is defined in the matrix form as
\begin{align}
    \boldsymbol{P}&=2\boldsymbol{G^\dagger K^{-1}G}\label{Pdef}.
\end{align}
where $k_{kk'}=\kappa_k\delta_{kk'}$.
This operator $\boldsymbol{P}$ represents the dissipation of the mechanical modes induced by the all the optical modes, which can be decomposed as 
\begin{align}
    \boldsymbol{P}&=\sum_{k=1}^M\boldsymbol{P}^{(k)}\label{PP},\\
    p_{jj'}^{(k)}&=2\frac{g_{k,j}^*g_{k,j'}}{\kappa_k}\label{p}.
\end{align}
Which can be viewd as dissipation induced by $k^\text{th}$ optical mode and only depends on driving strength $\Gamma_k$. Each $\boldsymbol{P}^{(k)}$ is proportional to the projection operator along  $\vec{g}^*_k=(g_{k,1}^*,\ g_{k,2}^*,\ ...,\ g_{k,N}^*)$. 
As a result, $\boldsymbol{P}^{(k)}$ cannot cool down any $\tilde{b}_j$ with coefficient vector $\vec{e}_{j}=(e_{j,1},\ e_{j,2},\ ...e_{j,N})^{T}$ orthogonal to $\vec{g}_k$.

By substituting Eq. (\ref{Pdef}) into Eq. (\ref{langevin_mec}), the Langevin equations of $b_j$ is rewritten as
\begin{align}
    \frac{\text{d} b_j}{\text{d}t}=(&-i\omega_{j}-\frac{\gamma_{j}}{2}) b_j-\sum_{j'=1}^Np_{jj'} b_{j'}-\sqrt{\gamma_{j}}b^\text{in}_j(t).
\end{align}
The corresponding effective mechanical Hamiltonian can thus be defined as
\begin{align}
    H_\text{eff}&=H_0-iD=\sum_{j=1}^N\hbar\omega_{j} b_j^\dagger b_j-i\sum_{j=1}^N\sum_{j'=1}^N b_j^\dagger p_{jj'} b_{j'},
\end{align}
with the master equation \cite{Praxmeyer_2019} given by
\begin{align}
    \frac{\text{d}\rho}{\text{d}t}\!=&\frac{i}{\hbar}[\rho,H_0]-\frac{1}{\hbar}[D,\rho]_++\frac{2}{\hbar}\rho \text{Tr}{[\rho D]}
    \nonumber\\
    &+\!\sum_{j=1}^N\frac{\gamma_{j}(n_\text{th}\!+\!1)}{2}[2  b_j\rho  b_j^\dagger- b_j^\dagger  b_j\rho-\rho  b_j^\dagger  b_j]\nonumber\\
    &+\!\sum_{j=1}^N\frac{\gamma_{j}n_\text{th}}{2}[2  b_j^\dagger\rho  b_j- b_j  b_j^\dagger\rho-\rho  b_j  b_j^\dagger].
\end{align}
Here, the steady state phonon can be calculated as $\langle  b_j^\dagger  b_{j'}\rangle=\text{Tr}[\rho b_j^\dagger  b_{j'}]$. 
The nonlinear terms $\langle  b_j^\dagger  b_{j'}^\dagger b_{j''}  b_{j'''}\rangle$ and $\langle  b_j^\dagger  b_{j'}\rangle\langle  b_{j''}^\dagger  b_{j'''}\rangle$ are expected to be very small ($\langle  b_j^\dagger  b_{j'}\rangle\ll n_\text{th}$) and can be ignored, which simplifies the Lyapunov equation to
\begin{align}
    &\frac{\text{d}}{\text{d}t}\langle  b_{j'}^\dagger  b_j\rangle=(i\omega_{j'}-i\omega_{j}-\frac{\gamma_j+\gamma_{j'}}{2})\langle  b_{j'}^\dagger  b_j\rangle\nonumber\\ &-\sum_{j''=1}^N[p_{jj''}\langle  b_{j'}^\dagger  b_{j''}\rangle+p_{j'j''}\langle  b_{j''}^\dagger  b_j\rangle]
    +\gamma_{j} \delta_{j'j}n_\text{th}.
\end{align}
In the case of $M=N=2$ and with symmetric parameters $\kappa_1=\kappa_2=\kappa$, $\gamma_1=\gamma_2=\gamma$, and $\Gamma_1=\Gamma_2=\Gamma$ assumed, the steady-state total phonon number reads 
\begin{align}
    &n^{(s)}_\text{tot}=\frac{2\gamma n_\text{th}(\gamma+2\Gamma)}{(\gamma+2\Gamma)^2-4\Gamma^2\cos^2\theta\frac{(\gamma+2\Gamma)^2}{(\gamma+2\Gamma)^2+\delta\omega_\text{mec}^2}}\label{adresult},
\end{align}
in which $\delta\omega_\text{mec}=|\omega_1-\omega_2|$, $\theta$ denotes the cross angle between the coupling vectors $\vec{g}_1$ and $\vec{g}_2$ with $0\leq\theta\leq\pi/2$.

With $\delta\omega_\text{mec}\ll\Gamma$, the Eq. (\ref{adresult}) is simplified to Eq. (\ref{align*:ntot}) of the main text.

\begin{figure*}[htbp]
\centering
\includegraphics[]{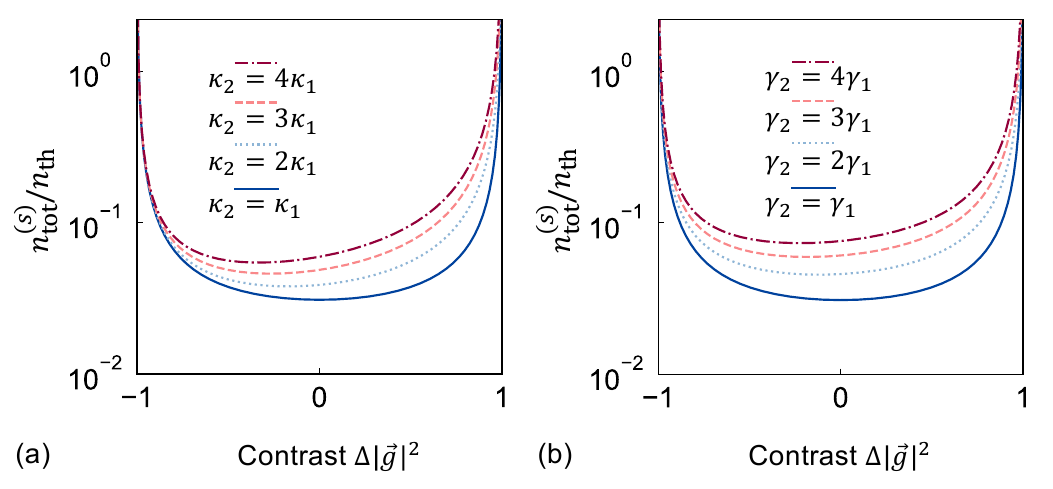}
\caption{(a) Total steady-state phonon number $n^{(s)}_\text{tot}$ as a function of contrast $\Delta |\vec{g}|^2=(|\vec{g}_1|^2-|\vec{g}_2|^2)/(|\vec{g}_1|^2+|\vec{g}_2|^2)$ with different $\kappa_2/\kappa_1$, where $|\vec{g}_1|^2+|\vec{g}_2|^2$ is kept constant. (b) Total steady-state phonon number $n^{s}_\text{tot}$ as a function of contrast $\Delta |\vec{g}|^2$ with different $\gamma_2/\gamma_1$, where $|\vec{g}_1|^2+|\vec{g}_2|^2$ is kept constant.
Here $\bar{\omega}_\text{mec}=20\kappa_1$, $\gamma_1=0.0001\kappa_1$, $\theta=\pi/4$, $|\vec{g}_1|^2+|\vec{g}_2|^2=0.5\kappa_1^2$. Rest parameters are same as Fig. 1.} 
\label{figSI2}
\end{figure*}

Without the symmetric-parameter assumption, the steady-state total phonon number with different $\kappa_1/\kappa_2$ and $\gamma_1/\gamma_2$, as a function of the drive contrast $\Delta |\vec{g}|^2=(|\vec{g}_1|^2-|\vec{g}_2|^2)/(|\vec{g}_1|^2+|\vec{g}_2|^2)$ is numerically calculated and presented in Fig. \ref{figSI2}.
It can be seen that the cooling performance is not significantly affected, except that the best cooling condition is reached when the two optical modes are unevenly pumped.
The analysis is then extended to systems with $N>2$, in which the coupling vectors are chosen to be linearly independent.
{\textcolor{black}{When the number of optical modes $M<N$, non zero dark modes exist. 
As the phonon numbers of the dark modes are approximately $n_\text{th}$ and far outweigh that of the bright modes, the total phonon number $n^{(s)}_\text{tot}\approx(N-M)n_\text{th}$.}} 

{\textcolor{black}{When $M\geq N$, all dark modes are eliminated and the above approximation fails. 
As $M$ exceeds $N$, the new optical modes inevitably have linear dependent coupling vectors with the existing $N$ optical modes and  thus can be understood as solely the increase of their drive strengths. Therefore, the following calculations focus on the $M=N$ case. }}
By assuming $\gamma_{j}=\gamma$, and $\gamma,\delta\omega_\text{mec}\ll \Gamma_k$, the Lyapunov equation Eq. (\ref{equ:1}-\ref{equ:3'}) becomes 
\begin{align}\label{pv}
    \boldsymbol{P^T V}+\boldsymbol{VP^T}=\gamma n_\text{th}\boldsymbol{I},
\end{align}
where $v_{j'j}=\langle  b_{j'}^\dagger  b_j\rangle$ and $n^{(s)}_\text{tot}=\text{Tr}\boldsymbol{V}$. 
As $\boldsymbol{P}$ is Hermitian, it is unitarily diagonalizable as $\boldsymbol{P^T=S^\dagger \Lambda S}$, with diagonal elements $\lambda_{j'j}=\lambda_{j}\delta_{j'j}$ and the unitary matrix $\boldsymbol{S}$ satisfying $\boldsymbol{S^\dagger S}=I$.
After the substitution, Eq. (\ref{pv}) can be rewritten as
\begin{align}
    [\boldsymbol{SVS^\dagger}]_{j'j}&=\frac{\gamma n_\text{th}\delta_{j'j}}{2\lambda_{j}},\\
    n^{(s)}_\text{tot}&=\text{Tr}\boldsymbol{V}
    =\frac{\gamma n_\text{th}}{2}\sum_{j=1}^N\lambda_{j}^{-1}
    =\frac{\gamma n_\text{th}}{2}\text{Tr}[\boldsymbol{P}^{-1}]\label{generalntot}.
\end{align}
Substituting Eq. (\ref{Pdef})
\begin{align}
    \text{Tr}[\boldsymbol{P}^{-1}]
    &=\frac{\kappa}{2}\parallel \boldsymbol{G}^{-1}\parallel_2,\\
    n^{(s)}_\text{tot}&=\frac{\kappa}{4}\gamma\kappa n_\text{th}\parallel \boldsymbol{G}^{-1}\parallel_2.
\end{align}
The parameter $\theta_k$ is defined as cross angle between $\vec{g}_k$ and the hyper surface spanned by the rest coupling vectors, which can be expressed as
\begin{align}
    \theta_k=\arcsin\frac{\vec{g}_k^*\cdot\vec{c_k}}{|\vec{g}_k||\vec{c}_k|}
\end{align}
where $\vec{c}_k$ is reciprocal vector of $\vec{g}_k$, satisfying $\vec{g}_k^*\cdot\vec{c}_j=\delta_{kj}$ and $0\leq\theta_k\leq\pi/2$.
Vectors $\vec {c}_k$ can be organized into a matrix $\boldsymbol{C}=(\boldsymbol{G}^{\dagger})^{-1}$, and by substituting $\boldsymbol{C}$ into Eq. (\ref{generalntot}), it is derived that
\begin{align}
     n^{(s)}_\text{tot}&=\frac{\gamma n_\text{th}}{4}\text{Tr} \boldsymbol{G}^{-1}\boldsymbol{KG}^{\dagger -1}=\frac{\gamma n_\text{th}}{4}\text{Tr} \boldsymbol{C}^\dagger \boldsymbol{KC}\nonumber\\
     &=\frac{\gamma n_\text{th}}{4}\sum_{k=1}^N\kappa_{k}|\vec{c}_k|^2
    =\frac{\gamma n_\text{th}}{4}\sum_{k=1}^N\frac{1}{\Gamma_{k}\sin^2\theta_{k}},
\end{align}
where $\Gamma_{k}=|\vec{g}_k|^2/\kappa_k$.

\section{Cooling simulation in a realistic optomechanical system}\label{C} 
The system investigated in Fig. \ref{fig4}(a) of the main text is composed of a silicon nitride membrane inserted into a Fabry-P$\acute{\text{e}}$rot optical cavity with fixed end mirrors, as schematically presented in Fig. \ref{figSI4}.
The designed system configuration and all the system parameters are based on the experimental works \cite{Purdy_2012,Jayich_2008,PhysRevLett.116.147202}, and the theoretical framework to describe the mechanical modes and optomechanical interactions is reported in Ref. \cite{PhysRevLett.108.083603}. 
\begin{figure}[h]
\includegraphics[width=8cm]{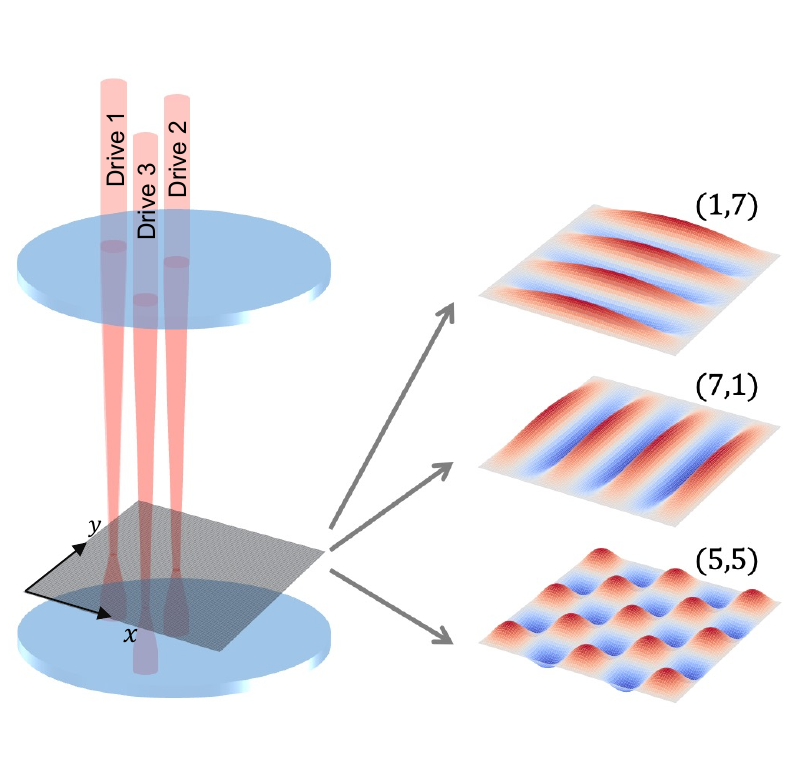}
\caption{Membrane-based setup for muti-mode cooling simulation performed in Fig. 4(a).} 
\label{figSI4}
\end{figure}

The membrane is assumed to be in square shape with edge length $l=\SI{1}{\milli\metre}$ and thickness $h=\SI{40}{nm}$. 
The mass density, Young's modulus $E$, and the Poisson's ratio of silicon nitride is taken as $\rho=\SI{2.7}{g/cm^3}$, $E=(200+0.01\text{i})\, \text{GPa}$, and $\nu=0.25$, respectively. 
The membrane is clamped on its four edges with a tensile stress of $\sigma=\SI{0.3}{\giga\pascal}$.
The strain field of the $(m,n)^\text{th}$ order mechanical drum mode can be written as \cite{PhysRevLett.108.083603}
\begin{align}
    W_{m,n}(x,y)=A_{m,n}\sin\frac{n\pi x}{l}\sin\frac{m\pi y}{l},
\end{align}
in which $A_{m,n}$ represents the peak strain amplitude of the $(m,n)^\text{th}$ mode.
The corresponding mode frequency $\omega_{m,n}$ and damping rate $\gamma_{m,n}$ are
\begin{align}
    \omega_{m,n}&=\frac{\pi}{l}\sqrt{\frac{\sigma(m^2+n^2)}{\rho}},\label{mec_freq}\\
    \gamma_{m,n}&=s\epsilon[1+\frac{\pi^2(m^2+n^2)}{4}]\omega_{m,n},\label{mec_linewidth}
\end{align}
respectively, where 
\begin{equation}
    \epsilon=\frac{h}{l}\sqrt{\frac{E}{3\sigma(1-\nu^2)}}.
\end{equation}
In Fig. 4(a) of the main text, three-fold degenerate modes, $(1,7)$, $(7,1)$, and $(5,5)$ are investigated, with degenerate frequencies and linewidths of $\omega/2\pi=\SI{1.178}{MHz}$ and $\gamma/2\pi=\SI{39.0}{mHz}$, according to Eq. (\ref{mec_freq}) and \ref{mec_linewidth}.

The optical cavity is assumed to possess a length of $L=\SI{6}{mm}$ and the fineness of $F=2.58\times 10^4$. 
The frequency of the optical modes is chosen as $\omega_c/2\pi=2.817\times 10^{16}\SI{}{Hz}$ ($\SI{1064}{nm}$), and thus the optical linewidth is calculated to be $\kappa/2\pi= c/(2FL)=\SI{0.967}{MHz}$.
The transverse optical modes are assumed to be Gaussian modes centered at different positions of the membrane, with the light intensity on the membrane plane expressed as
\begin{align}
    I(x,y)=\frac{\exp[(x-x_0)^2/d^2+(y-y_0)^2/d^2]}{\pi d^2},
\end{align}
where $d=\SI{90}{\micro \metre}$ is the waist radius assumed for the Gaussian mode, and  $(x_0,y_0)$ is the center of the transverse mode profile.

The single photon optomechanical coupling strength is calculated by the transverse mode overlap between the mechanical and optical modes as \cite{Purdy_2012}
\begin{align}
    g^S_{(m,n)}(I)&=x^\text{ZPF}_{m,n}\frac{d\omega_c}{dL}\eta_{m,n}(I),
\end{align}
where the zero point fluctuation $x^\text{ZPF}_{m,n}=[2\hbar/(\rho hl^2\omega_{m,n})]^{1/2}=5.13\times 10^{-16}\ \SI{}{m}$, the frequency shift coefficient $d\omega_c/dL=2\pi c/\lambda L=2.95\times 10^{17}\ \SI{}{Hz/m}$, and the overlap integral
\begin{align}
    \eta_{m,n}(I)&=\iint\displaylimits_{x,y\in[0,l]}\!\!\left[ I(x,y)W_{m,n}(x,y)/A_{m,n}\right] dxdy.
\end{align}
As shown in Fig. 4(a) of the main text, the centers of three driven optical modes are assumed to
\begin{align}
    (x_{01},y_{01})=(\frac{2}{7}l,\frac{1}{2}l)\ \ \ \text{for Drive 1},\\
    (x_{02},y_{02})=(\frac{1}{2}l,\frac{1}{2}l)\ \ \ \text{for Drive 2},\\
    (x_{03},y_{03})=(\frac{1}{2}l,\frac{2}{7}l)\ \ \ \text{for Drive 3},
\end{align}
and the single photon coupling strengths between different optical and mechanical modes are listed in Table \ref{table} (unit: $\SI{}{Hz}$). {\textcolor{black}{For an optical mode coupled with multiple near-degenerate mechanical modes, the length of the single photon coupling vector is proportion to $x^\text{ZPF}_{m,n}\frac{d\omega_c}{dL}$, while the direction of it is dominated by the overlap integrals $\eta_{m,n}(I)$, which is possible to be controlled by the spatial profile of the optical modes in many systems. 
In our simulation, specifically, the centers of the Gaussian beams are approximately focused on the nodes or antinodes of the mechanical modes, to minimize or maximize the optomechanical coupling strength, which thus gives rise to large $\theta$s.}} 
As a comparison with real systems, the single photon coupling strength reported in \cite{Purdy_2012} is $\SI{105.23}{Hz}$, which is comparable to our calculation.
In Fig. 4(a), the driving strength $\Gamma$ of each optical mode increases from $0$ to $\Gamma_{0}=4000\gamma=\SI{980}{Hz}$, corresponding to a maximum intracavity photon number $n_\text{opt}=1.95\times 10^{5}$, which is comparable to the value $n_\text{opt}=3\times 10^{5}$ reported in \cite{Purdy_2012}. 

\begin{table}[!h]
\setlength{\tabcolsep}{6mm}
\renewcommand\arraystretch{2}
\begin{center}
\begin{tabular}{|c|c|c|c|}
    \hline
     & $W_{1,7}$ & $W_{7,1}$ & $W_{5,5}$\\
    \hline
    $(\frac{2}{7}l,\frac{1}{2}l)$&43.61&0&54.38\\
    \hline
    $(\frac{1}{2}l,\frac{1}{2}l)$&55.78& 55.78 &-55.78\\
    \hline
    $(\frac{1}{2}l,\frac{2}{7}l)$&0&43.61&54.38\\
    \hline
\end{tabular}
\caption{Calculated single phonon coupling strengths $g^S$ between the selected mechanical and optical modes.}
\label{table}
\end{center}
\end{table}
In Fig. 4(a) of the main text, hybrid mechanical modes $\tilde{b}_j=\sum_{j'=1}^N e_{jj'}b_{j'}$ are calculated via Schmidt orthogonalization on the coupling vectors as
\begin{align}
    \vec{e'}_1&=\vec{g}_1,\nonumber\\
    \vec{e'}_2&=\vec{g}_2-\frac{\vec{g}_2\cdot\vec{e'}_1}{\vec{e'}'_1\cdot\vec{e'}_1}\vec{e'}_1,\nonumber\\
    &...\nonumber\\
    \vec{e'}_j&=\vec{g}_j-\sum_{k<j}\frac{\vec{g}_j\cdot\vec{e'}_k}{\vec{e'}_k\cdot\vec{e'}_k}\vec{e'}_k,\nonumber\\
    &...\nonumber\\
    \vec{e'}_N&=\vec{g}_N-\sum_{k<N}\frac{\vec{g}_N\cdot\vec{e'}_k}{\vec{e'}_k\cdot\vec{e'}_k}\vec{e'}_k.\nonumber
\end{align}
All coefficient vectors are then normalized by  $\vec{e}_j=\vec{e'}_j/|\vec{e'}_j|$ to ensure each $\tilde{b}_j$ being standard annihilation operators. 
In this way, when $k$ optical drives are on, the $j^\text{th}$ mechanical mode $\tilde{b}_j$ remains dark if $k<j$
since $\vec{g}_k^*\cdot\vec{e}_j=0$,  until the $j^\text{th}$ optical drive is turned on.
{\textcolor{black}{As the linearized system has one unique steady state, the cooling results do not depend the path of parameter-variation. 
When lasers are turned on simultaneously, the final steady state phonon number is the same as the sequential drive scheme, as shown in Fig. S4. }}
\begin{figure*}[htp]
\centering
\includegraphics[]{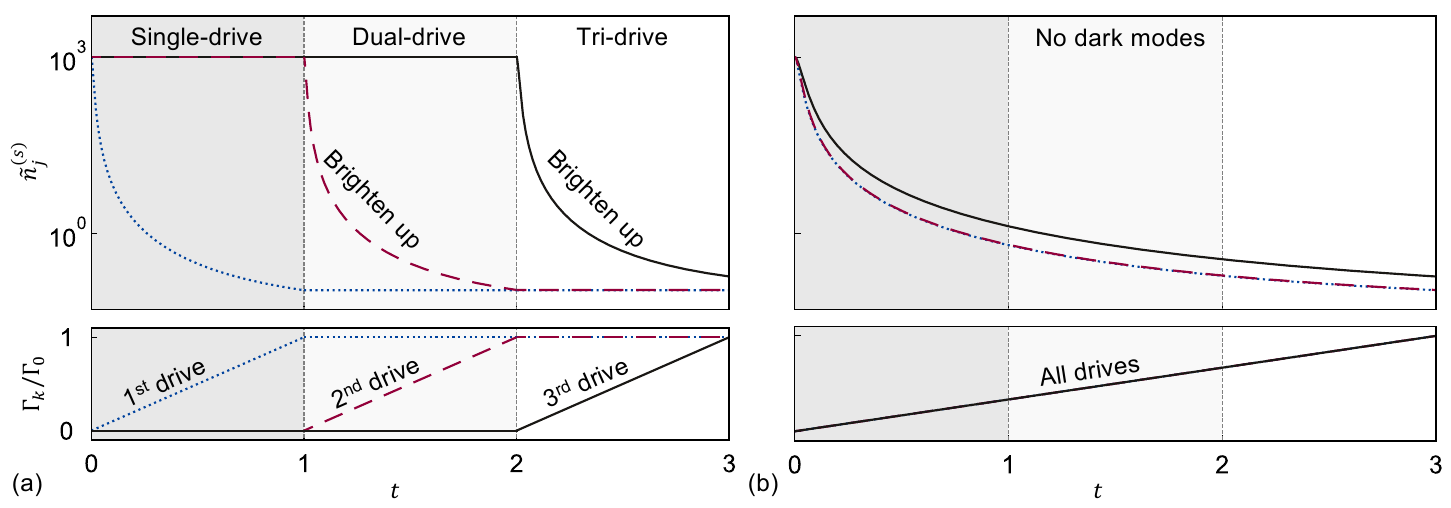}
\caption{{\textcolor{black}{Evolution of $n_\text{tot}^{(s)}$ of each mechanical eigenmode (upper panel), as the three optical drives are quasi-statically introduced (lower panel). (a): Successively turning on the drives. (b): Simultaneously turning on the drives. The parameters are the same as Fig. 4(a). }}} 
\label{figSI4}
\end{figure*}
In Fig. 4(b) of the main text, the single photon coupling strength is assumed to be $g^S_{k,j}=(0.6\delta_{k,j}+0.2)\times 10^{-3}\kappa$,{\textcolor{black}{$k\leq10;g^S_{11,j}=\delta_{1,j}\times 10^{-3}\kappa, g^S_{12,j}=\delta_{2,j}\times 10^{-3}\kappa$}}, with other parameters $\omega_j=\bar{\omega}_\text{mec}=20\kappa$, $\gamma_j=10^{-4}\kappa$, and $\Gamma_0=0.1\kappa$. 

\section{Calculation of the classical cooling limit}\label{D} 
The classical cooling limit is reached when the system enters the optomechanical strong coupling regime.
In this regime, the adiabatic elimination of the optical modes Eq. (\ref{no_thermal_opt}) no longer applies.
In the sideband-resolved regime, the classical cooling limit is obtained by solving Eq. (\ref{equ:1}), (\ref{equ:2}), and (\ref{equ:3}).

When $M=N=2$, the steady-state total phonon number of the two mechanical modes can be solved as
\begin{align}
    n^{(s)}_\text{tot}&=\frac{2Ln_\text{th}}{L+2s+4\sin^2\theta-\frac{4s^2 \cos^2\theta}{L +2s+4\sin^2\theta}},
\end{align}
where $L=\gamma[(s+2)^2-4\cos^2\theta]/\kappa$ and $s=(\gamma+\kappa)/\Gamma$.
Here, symmetric parameters $\delta\omega_\text{mec}=0$, $\gamma_1=\gamma_2=\gamma$, $\kappa_1=\kappa_2=\kappa$, $ \Gamma_1=\Gamma_2=\Gamma/2$ are assumed.

\hfill

\noindent
For any $\theta\neq 0$ and $s\approx0$
\begin{align}
    n^{(s)}_\text{tot}=\frac{2\gamma n_\text{th}}{\gamma+\kappa}.
\end{align}
If $\theta=0$, the same condition leads to
\begin{align}
    n^{(s)}_\text{tot}=(\frac{\gamma}{\gamma+\kappa}+1)n_\text{th}.
\end{align}

\section{Estimation of the quantum cooling limit}\label{E} 
The quantum cooling limit is calculated by the force noise power spectral density 
$S^{FF}_{j}(\omega)=\int_{-\infty}^\infty \langle F_j(t)F_j(0)\rangle e^{i\omega t}$. 
The optical force on the $j^\text{th}$ mechanical mode $F_j$ is written as
\begin{align}
    F_j=(x^\text{ZPF})^{-1}\sum_{k}^M (g_{k,j}a^\dagger_k+g^*_{k,j}a_k).
\end{align}
Substituting Eq. (\ref{eqn:a(omega)}) into $S^{FF}_{j}$ and only keeping the quantum noise input $a_k^\text{in}$, it is derived that
\begin{align}
    S^{FF}_{j}(\omega)=(x^\text{ZPF})^{-2}\sum_{k=1}^M\frac{|g_{k,j}|^2}{(\omega-\delta_k)^2+\frac{\kappa_k^2}{4}}.
\end{align}
The quantum cooling limit can be calculated as
\begin{align}
    &n^\text{Q}_j=\frac{S^{FF}_{j}(-\bar{\omega}_\text{mec})}{S^{FF}_{j}(\bar{\omega}_\text{mec})-S^{FF}_{j}(-\bar{\omega}_\text{mec})}.
\end{align}
Given $\delta_k=\bar{\omega}_\text{mec}$, $\kappa_k=\kappa$, this expression reduces to
\begin{align}
    n^\text{Q}_j=\frac{\kappa^2}{16\bar{\omega}_\text{mec}^2},
\end{align}
thus the quantum cooling limit does not depend on $\boldsymbol{G}$ under current approximation, and the result is the same as the single mode case.
In the resolved-sideband regime investigated in the main text with $\bar{\omega}_\text{mec}=20\kappa$, the quantum cooling limit of every mechanical mode $n^\text{Q}_j<10^{-3}$, which is negligible comparing with the classical cooling limit. 

\bibliography{arxiv}

\providecommand{\noopsort}[1]{}\providecommand{\singleletter}[1]{#1}%
\begin{thebibliography}{56}%
\makeatletter
\providecommand \@ifxundefined [1]{%
 \@ifx{#1\undefined}
}%
\providecommand \@ifnum [1]{%
 \ifnum #1\expandafter \@firstoftwo
 \else \expandafter \@secondoftwo
 \fi
}%
\providecommand \@ifx [1]{%
 \ifx #1\expandafter \@firstoftwo
 \else \expandafter \@secondoftwo
 \fi
}%
\providecommand \natexlab [1]{#1}%
\providecommand \enquote  [1]{``#1''}%
\providecommand \bibnamefont  [1]{#1}%
\providecommand \bibfnamefont [1]{#1}%
\providecommand \citenamefont [1]{#1}%
\providecommand \href@noop [0]{\@secondoftwo}%
\providecommand \href [0]{\begingroup \@sanitize@url \@href}%
\providecommand \@href[1]{\@@startlink{#1}\@@href}%
\providecommand \@@href[1]{\endgroup#1\@@endlink}%
\providecommand \@sanitize@url [0]{\catcode `\\12\catcode `\$12\catcode
  `\&12\catcode `\#12\catcode `\^12\catcode `\_12\catcode `\%12\relax}%
\providecommand \@@startlink[1]{}%
\providecommand \@@endlink[0]{}%
\providecommand \url  [0]{\begingroup\@sanitize@url \@url }%
\providecommand \@url [1]{\endgroup\@href {#1}{\urlprefix }}%
\providecommand \urlprefix  [0]{URL }%
\providecommand \Eprint [0]{\href }%
\providecommand \doibase [0]{https://doi.org/}%
\providecommand \selectlanguage [0]{\@gobble}%
\providecommand \bibinfo  [0]{\@secondoftwo}%
\providecommand \bibfield  [0]{\@secondoftwo}%
\providecommand \translation [1]{[#1]}%
\providecommand \BibitemOpen [0]{}%
\providecommand \bibitemStop [0]{}%
\providecommand \bibitemNoStop [0]{.\EOS\space}%
\providecommand \EOS [0]{\spacefactor3000\relax}%
\providecommand \BibitemShut  [1]{\csname bibitem#1\endcsname}%
\let\auto@bib@innerbib\@empty
\bibitem [{\citenamefont {Aspelmeyer}\ \emph {et~al.}(2014)\citenamefont
  {Aspelmeyer}, \citenamefont {Kippenberg},\ and\ \citenamefont
  {Marquardt}}]{Aspelmeyer2014}%
  \BibitemOpen
  \bibfield  {author} {\bibinfo {author} {\bibfnamefont {M.}~\bibnamefont
  {Aspelmeyer}}, \bibinfo {author} {\bibfnamefont {T.~J.}\ \bibnamefont
  {Kippenberg}},\ and\ \bibinfo {author} {\bibfnamefont {F.}~\bibnamefont
  {Marquardt}},\ }\bibfield  {title} {\bibinfo {title} {{Cavity
  optomechanics}},\ }\href {https://doi.org/10.1103/RevModPhys.86.1391}
  {\bibfield  {journal} {\bibinfo  {journal} {Rev. Mod. Phys.}\ }\textbf
  {\bibinfo {volume} {86}},\ \bibinfo {pages} {1391} (\bibinfo {year}
  {2014})}\BibitemShut {NoStop}%
\bibitem [{\citenamefont {Ockeloen-Korppi}\ \emph {et~al.}(2018)\citenamefont
  {Ockeloen-Korppi}, \citenamefont {Damsk{\"{a}}gg}, \citenamefont
  {Pirkkalainen}, \citenamefont {Asjad}, \citenamefont {Clerk}, \citenamefont
  {Massel}, \citenamefont {Woolley},\ and\ \citenamefont
  {Sillanp{\"{a}}{\"{a}}}}]{Ockeloen-Korppi2018}%
  \BibitemOpen
  \bibfield  {author} {\bibinfo {author} {\bibfnamefont {C.~F.}\ \bibnamefont
  {Ockeloen-Korppi}}, \bibinfo {author} {\bibfnamefont {E.}~\bibnamefont
  {Damsk{\"{a}}gg}}, \bibinfo {author} {\bibfnamefont {J.-M.}\ \bibnamefont
  {Pirkkalainen}}, \bibinfo {author} {\bibfnamefont {M.}~\bibnamefont {Asjad}},
  \bibinfo {author} {\bibfnamefont {A.~A.}\ \bibnamefont {Clerk}}, \bibinfo
  {author} {\bibfnamefont {F.}~\bibnamefont {Massel}}, \bibinfo {author}
  {\bibfnamefont {M.~J.}\ \bibnamefont {Woolley}},\ and\ \bibinfo {author}
  {\bibfnamefont {M.~A.}\ \bibnamefont {Sillanp{\"{a}}{\"{a}}}},\ }\bibfield
  {title} {\bibinfo {title} {{Stabilized entanglement of massive mechanical
  oscillators}},\ }\href {https://doi.org/10.1038/s41586-018-0038-x} {\bibfield
   {journal} {\bibinfo  {journal} {Nature}\ }\textbf {\bibinfo {volume}
  {556}},\ \bibinfo {pages} {478} (\bibinfo {year} {2018})}\BibitemShut
  {NoStop}%
\bibitem [{\citenamefont {Riedinger}\ \emph {et~al.}(2018)\citenamefont
  {Riedinger}, \citenamefont {Wallucks}, \citenamefont {Marinkovi{\'{c}}},
  \citenamefont {L{\"o}schnauer}, \citenamefont {Aspelmeyer}, \citenamefont
  {Hong},\ and\ \citenamefont {Gr{\"o}blacher}}]{Riedinger2018}%
  \BibitemOpen
  \bibfield  {author} {\bibinfo {author} {\bibfnamefont {R.}~\bibnamefont
  {Riedinger}}, \bibinfo {author} {\bibfnamefont {A.}~\bibnamefont {Wallucks}},
  \bibinfo {author} {\bibfnamefont {I.}~\bibnamefont {Marinkovi{\'{c}}}},
  \bibinfo {author} {\bibfnamefont {C.}~\bibnamefont {L{\"o}schnauer}},
  \bibinfo {author} {\bibfnamefont {M.}~\bibnamefont {Aspelmeyer}}, \bibinfo
  {author} {\bibfnamefont {S.}~\bibnamefont {Hong}},\ and\ \bibinfo {author}
  {\bibfnamefont {S.}~\bibnamefont {Gr{\"o}blacher}},\ }\bibfield  {title}
  {\bibinfo {title} {Remote quantum entanglement between two micromechanical
  oscillators},\ }\href {https://doi.org/10.1038/s41586-018-0036-z} {\bibfield
  {journal} {\bibinfo  {journal} {Nature}\ }\textbf {\bibinfo {volume} {556}},\
  \bibinfo {pages} {473} (\bibinfo {year} {2018})}\BibitemShut {NoStop}%
\bibitem [{\citenamefont {Ockeloen-Korppi}\ \emph {et~al.}(2016)\citenamefont
  {Ockeloen-Korppi}, \citenamefont {Damsk{\"{a}}gg}, \citenamefont
  {Pirkkalainen}, \citenamefont {Clerk}, \citenamefont {Woolley},\ and\
  \citenamefont {Sillanp{\"{a}}{\"{a}}}}]{Ockeloen-Korppi2016}%
  \BibitemOpen
  \bibfield  {author} {\bibinfo {author} {\bibfnamefont {C.~F.}\ \bibnamefont
  {Ockeloen-Korppi}}, \bibinfo {author} {\bibfnamefont {E.}~\bibnamefont
  {Damsk{\"{a}}gg}}, \bibinfo {author} {\bibfnamefont {J.-M.}\ \bibnamefont
  {Pirkkalainen}}, \bibinfo {author} {\bibfnamefont {A.~A.}\ \bibnamefont
  {Clerk}}, \bibinfo {author} {\bibfnamefont {M.~J.}\ \bibnamefont {Woolley}},\
  and\ \bibinfo {author} {\bibfnamefont {M.~A.}\ \bibnamefont
  {Sillanp{\"{a}}{\"{a}}}},\ }\bibfield  {title} {\bibinfo {title} {{Quantum
  Backaction Evading Measurement of Collective Mechanical Modes}},\ }\href
  {https://doi.org/10.1103/PhysRevLett.117.140401} {\bibfield  {journal}
  {\bibinfo  {journal} {Phys. Rev. Lett.}\ }\textbf {\bibinfo {volume} {117}},\
  \bibinfo {pages} {140401} (\bibinfo {year} {2016})}\BibitemShut {NoStop}%
\bibitem [{\citenamefont {Massel}\ \emph {et~al.}(2012)\citenamefont {Massel},
  \citenamefont {Cho}, \citenamefont {Pirkkalainen}, \citenamefont {Hakonen},
  \citenamefont {Heikkil{\"{a}}},\ and\ \citenamefont
  {Sillanp{\"{a}}{\"{a}}}}]{Massel2012}%
  \BibitemOpen
  \bibfield  {author} {\bibinfo {author} {\bibfnamefont {F.}~\bibnamefont
  {Massel}}, \bibinfo {author} {\bibfnamefont {S.~U.}\ \bibnamefont {Cho}},
  \bibinfo {author} {\bibfnamefont {J.-M.}\ \bibnamefont {Pirkkalainen}},
  \bibinfo {author} {\bibfnamefont {P.~J.}\ \bibnamefont {Hakonen}}, \bibinfo
  {author} {\bibfnamefont {T.~T.}\ \bibnamefont {Heikkil{\"{a}}}},\ and\
  \bibinfo {author} {\bibfnamefont {M.~A.}\ \bibnamefont
  {Sillanp{\"{a}}{\"{a}}}},\ }\bibfield  {title} {\bibinfo {title} {{Multimode
  circuit optomechanics near the quantum limit}},\ }\href
  {https://doi.org/10.1038/ncomms1993} {\bibfield  {journal} {\bibinfo
  {journal} {Nat. Commun.}\ }\textbf {\bibinfo {volume} {3}},\ \bibinfo {pages}
  {987} (\bibinfo {year} {2012})}\BibitemShut {NoStop}%
\bibitem [{\citenamefont {Liao}\ and\ \citenamefont
  {Tian}(2016)}]{PhysRevLett.116.163602}%
  \BibitemOpen
  \bibfield  {author} {\bibinfo {author} {\bibfnamefont {J.-Q.}\ \bibnamefont
  {Liao}}\ and\ \bibinfo {author} {\bibfnamefont {L.}~\bibnamefont {Tian}},\
  }\bibfield  {title} {\bibinfo {title} {Macroscopic quantum superposition in
  cavity optomechanics},\ }\href
  {https://doi.org/10.1103/PhysRevLett.116.163602} {\bibfield  {journal}
  {\bibinfo  {journal} {Phys. Rev. Lett.}\ }\textbf {\bibinfo {volume} {116}},\
  \bibinfo {pages} {163602} (\bibinfo {year} {2016})}\BibitemShut {NoStop}%
\bibitem [{\citenamefont {Pepper}\ \emph {et~al.}(2012)\citenamefont {Pepper},
  \citenamefont {Ghobadi}, \citenamefont {Jeffrey}, \citenamefont {Simon},\
  and\ \citenamefont {Bouwmeester}}]{Pepper2012}%
  \BibitemOpen
  \bibfield  {author} {\bibinfo {author} {\bibfnamefont {B.}~\bibnamefont
  {Pepper}}, \bibinfo {author} {\bibfnamefont {R.}~\bibnamefont {Ghobadi}},
  \bibinfo {author} {\bibfnamefont {E.}~\bibnamefont {Jeffrey}}, \bibinfo
  {author} {\bibfnamefont {C.}~\bibnamefont {Simon}},\ and\ \bibinfo {author}
  {\bibfnamefont {D.}~\bibnamefont {Bouwmeester}},\ }\bibfield  {title}
  {\bibinfo {title} {{Optomechanical Superpositions via Nested
  Interferometry}},\ }\href {https://doi.org/10.1103/PhysRevLett.109.023601}
  {\bibfield  {journal} {\bibinfo  {journal} {Phys. Rev. Lett.}\ }\textbf
  {\bibinfo {volume} {109}},\ \bibinfo {pages} {023601} (\bibinfo {year}
  {2012})}\BibitemShut {NoStop}%
\bibitem [{\citenamefont {Liao}\ \emph {et~al.}(2014)\citenamefont {Liao},
  \citenamefont {Wu},\ and\ \citenamefont {Nori}}]{PhysRevA.89.014302}%
  \BibitemOpen
  \bibfield  {author} {\bibinfo {author} {\bibfnamefont {J.-Q.}\ \bibnamefont
  {Liao}}, \bibinfo {author} {\bibfnamefont {Q.-Q.}\ \bibnamefont {Wu}},\ and\
  \bibinfo {author} {\bibfnamefont {F.}~\bibnamefont {Nori}},\ }\bibfield
  {title} {\bibinfo {title} {Entangling two macroscopic mechanical mirrors in a
  two-cavity optomechanical system},\ }\href
  {https://doi.org/10.1103/PhysRevA.89.014302} {\bibfield  {journal} {\bibinfo
  {journal} {Phys. Rev. A}\ }\textbf {\bibinfo {volume} {89}},\ \bibinfo
  {pages} {014302} (\bibinfo {year} {2014})}\BibitemShut {NoStop}%
\bibitem [{\citenamefont {Mancini}\ \emph {et~al.}(2002)\citenamefont
  {Mancini}, \citenamefont {Giovannetti}, \citenamefont {Vitali},\ and\
  \citenamefont {Tombesi}}]{PhysRevLett.88.120401}%
  \BibitemOpen
  \bibfield  {author} {\bibinfo {author} {\bibfnamefont {S.}~\bibnamefont
  {Mancini}}, \bibinfo {author} {\bibfnamefont {V.}~\bibnamefont
  {Giovannetti}}, \bibinfo {author} {\bibfnamefont {D.}~\bibnamefont
  {Vitali}},\ and\ \bibinfo {author} {\bibfnamefont {P.}~\bibnamefont
  {Tombesi}},\ }\bibfield  {title} {\bibinfo {title} {Entangling macroscopic
  oscillators exploiting radiation pressure},\ }\href
  {https://doi.org/10.1103/PhysRevLett.88.120401} {\bibfield  {journal}
  {\bibinfo  {journal} {Phys. Rev. Lett.}\ }\textbf {\bibinfo {volume} {88}},\
  \bibinfo {pages} {120401} (\bibinfo {year} {2002})}\BibitemShut {NoStop}%
\bibitem [{\citenamefont {Wei}\ \emph {et~al.}(2006)\citenamefont {Wei},
  \citenamefont {Liu}, \citenamefont {Sun},\ and\ \citenamefont
  {Nori}}]{PhysRevLett.97.237201}%
  \BibitemOpen
  \bibfield  {author} {\bibinfo {author} {\bibfnamefont {L.~F.}\ \bibnamefont
  {Wei}}, \bibinfo {author} {\bibfnamefont {Y.-x.}\ \bibnamefont {Liu}},
  \bibinfo {author} {\bibfnamefont {C.~P.}\ \bibnamefont {Sun}},\ and\ \bibinfo
  {author} {\bibfnamefont {F.}~\bibnamefont {Nori}},\ }\bibfield  {title}
  {\bibinfo {title} {Probing tiny motions of nanomechanical resonators:
  Classical or quantum mechanical?},\ }\href
  {https://doi.org/10.1103/PhysRevLett.97.237201} {\bibfield  {journal}
  {\bibinfo  {journal} {Phys. Rev. Lett.}\ }\textbf {\bibinfo {volume} {97}},\
  \bibinfo {pages} {237201} (\bibinfo {year} {2006})}\BibitemShut {NoStop}%
\bibitem [{\citenamefont {Poot}\ and\ \citenamefont {{van der
  Zant}}(2012)}]{POOT2012273}%
  \BibitemOpen
  \bibfield  {author} {\bibinfo {author} {\bibfnamefont {M.}~\bibnamefont
  {Poot}}\ and\ \bibinfo {author} {\bibfnamefont {H.~S.}\ \bibnamefont {{van
  der Zant}}},\ }\bibfield  {title} {\bibinfo {title} {Mechanical systems in
  the quantum regime},\ }\href
  {https://doi.org/https://doi.org/10.1016/j.physrep.2011.12.004} {\bibfield
  {journal} {\bibinfo  {journal} {Phys. Rep.}\ }\textbf {\bibinfo {volume}
  {511}},\ \bibinfo {pages} {273} (\bibinfo {year} {2012})},\ \bibinfo {note}
  {mechanical systems in the quantum regime}\BibitemShut {NoStop}%
\bibitem [{\citenamefont {Liu}\ \emph {et~al.}(2020)\citenamefont {Liu},
  \citenamefont {Pagliano}, \citenamefont {van Veldhoven}, \citenamefont
  {Pogoretskiy}, \citenamefont {Jiao},\ and\ \citenamefont {Fiore}}]{Liu2020}%
  \BibitemOpen
  \bibfield  {author} {\bibinfo {author} {\bibfnamefont {T.}~\bibnamefont
  {Liu}}, \bibinfo {author} {\bibfnamefont {F.}~\bibnamefont {Pagliano}},
  \bibinfo {author} {\bibfnamefont {R.}~\bibnamefont {van Veldhoven}}, \bibinfo
  {author} {\bibfnamefont {V.}~\bibnamefont {Pogoretskiy}}, \bibinfo {author}
  {\bibfnamefont {Y.}~\bibnamefont {Jiao}},\ and\ \bibinfo {author}
  {\bibfnamefont {A.}~\bibnamefont {Fiore}},\ }\bibfield  {title} {\bibinfo
  {title} {Integrated nano-optomechanical displacement sensor with ultrawide
  optical bandwidth},\ }\href {https://doi.org/10.1038/s41467-020-16269-7}
  {\bibfield  {journal} {\bibinfo  {journal} {Nat. Commun.}\ }\textbf {\bibinfo
  {volume} {11}},\ \bibinfo {pages} {2407} (\bibinfo {year}
  {2020})}\BibitemShut {NoStop}%
\bibitem [{\citenamefont {Fogliano}\ \emph {et~al.}(2021)\citenamefont
  {Fogliano}, \citenamefont {Besga}, \citenamefont {Reigue}, \citenamefont
  {Mercier~de L{\'e}pinay}, \citenamefont {Heringlake}, \citenamefont
  {Gouriou}, \citenamefont {Eyraud}, \citenamefont {Wernsdorfer}, \citenamefont
  {Pigeau},\ and\ \citenamefont {Arcizet}}]{Fogliano2021}%
  \BibitemOpen
  \bibfield  {author} {\bibinfo {author} {\bibfnamefont {F.}~\bibnamefont
  {Fogliano}}, \bibinfo {author} {\bibfnamefont {B.}~\bibnamefont {Besga}},
  \bibinfo {author} {\bibfnamefont {A.}~\bibnamefont {Reigue}}, \bibinfo
  {author} {\bibfnamefont {L.}~\bibnamefont {Mercier~de L{\'e}pinay}}, \bibinfo
  {author} {\bibfnamefont {P.}~\bibnamefont {Heringlake}}, \bibinfo {author}
  {\bibfnamefont {C.}~\bibnamefont {Gouriou}}, \bibinfo {author} {\bibfnamefont
  {E.}~\bibnamefont {Eyraud}}, \bibinfo {author} {\bibfnamefont
  {W.}~\bibnamefont {Wernsdorfer}}, \bibinfo {author} {\bibfnamefont
  {B.}~\bibnamefont {Pigeau}},\ and\ \bibinfo {author} {\bibfnamefont
  {O.}~\bibnamefont {Arcizet}},\ }\bibfield  {title} {\bibinfo {title}
  {Ultrasensitive nano-optomechanical force sensor operated at dilution
  temperatures},\ }\href {https://doi.org/10.1038/s41467-021-24318-y}
  {\bibfield  {journal} {\bibinfo  {journal} {Nat. Commun.}\ }\textbf {\bibinfo
  {volume} {12}},\ \bibinfo {pages} {4124} (\bibinfo {year}
  {2021})}\BibitemShut {NoStop}%
\bibitem [{\citenamefont {Krause}\ \emph {et~al.}(2012)\citenamefont {Krause},
  \citenamefont {Winger}, \citenamefont {Blasius}, \citenamefont {Lin},\ and\
  \citenamefont {Painter}}]{Krause2012}%
  \BibitemOpen
  \bibfield  {author} {\bibinfo {author} {\bibfnamefont {A.~G.}\ \bibnamefont
  {Krause}}, \bibinfo {author} {\bibfnamefont {M.}~\bibnamefont {Winger}},
  \bibinfo {author} {\bibfnamefont {T.~D.}\ \bibnamefont {Blasius}}, \bibinfo
  {author} {\bibfnamefont {Q.}~\bibnamefont {Lin}},\ and\ \bibinfo {author}
  {\bibfnamefont {O.}~\bibnamefont {Painter}},\ }\bibfield  {title} {\bibinfo
  {title} {A high-resolution microchip optomechanical accelerometer},\ }\href
  {https://doi.org/10.1038/nphoton.2012.245} {\bibfield  {journal} {\bibinfo
  {journal} {Nat. Photonics}\ }\textbf {\bibinfo {volume} {6}},\ \bibinfo
  {pages} {768} (\bibinfo {year} {2012})}\BibitemShut {NoStop}%
\bibitem [{\citenamefont {Mancini}\ \emph {et~al.}(1998)\citenamefont
  {Mancini}, \citenamefont {Vitali},\ and\ \citenamefont
  {Tombesi}}]{PhysRevLett.80.688}%
  \BibitemOpen
  \bibfield  {author} {\bibinfo {author} {\bibfnamefont {S.}~\bibnamefont
  {Mancini}}, \bibinfo {author} {\bibfnamefont {D.}~\bibnamefont {Vitali}},\
  and\ \bibinfo {author} {\bibfnamefont {P.}~\bibnamefont {Tombesi}},\
  }\bibfield  {title} {\bibinfo {title} {Optomechanical cooling of a
  macroscopic oscillator by homodyne feedback},\ }\href
  {https://doi.org/10.1103/PhysRevLett.80.688} {\bibfield  {journal} {\bibinfo
  {journal} {Phys. Rev. Lett.}\ }\textbf {\bibinfo {volume} {80}},\ \bibinfo
  {pages} {688} (\bibinfo {year} {1998})}\BibitemShut {NoStop}%
\bibitem [{\citenamefont {Wilson-Rae}\ \emph {et~al.}(2007)\citenamefont
  {Wilson-Rae}, \citenamefont {Nooshi}, \citenamefont {Zwerger},\ and\
  \citenamefont {Kippenberg}}]{PhysRevLett.99.093901}%
  \BibitemOpen
  \bibfield  {author} {\bibinfo {author} {\bibfnamefont {I.}~\bibnamefont
  {Wilson-Rae}}, \bibinfo {author} {\bibfnamefont {N.}~\bibnamefont {Nooshi}},
  \bibinfo {author} {\bibfnamefont {W.}~\bibnamefont {Zwerger}},\ and\ \bibinfo
  {author} {\bibfnamefont {T.~J.}\ \bibnamefont {Kippenberg}},\ }\bibfield
  {title} {\bibinfo {title} {Theory of ground state cooling of a mechanical
  oscillator using dynamical backaction},\ }\href
  {https://doi.org/10.1103/PhysRevLett.99.093901} {\bibfield  {journal}
  {\bibinfo  {journal} {Phys. Rev. Lett.}\ }\textbf {\bibinfo {volume} {99}},\
  \bibinfo {pages} {093901} (\bibinfo {year} {2007})}\BibitemShut {NoStop}%
\bibitem [{\citenamefont {Marquardt}\ \emph {et~al.}(2007)\citenamefont
  {Marquardt}, \citenamefont {Chen}, \citenamefont {Clerk},\ and\ \citenamefont
  {Girvin}}]{PhysRevLett.99.093902}%
  \BibitemOpen
  \bibfield  {author} {\bibinfo {author} {\bibfnamefont {F.}~\bibnamefont
  {Marquardt}}, \bibinfo {author} {\bibfnamefont {J.~P.}\ \bibnamefont {Chen}},
  \bibinfo {author} {\bibfnamefont {A.~A.}\ \bibnamefont {Clerk}},\ and\
  \bibinfo {author} {\bibfnamefont {S.~M.}\ \bibnamefont {Girvin}},\ }\bibfield
   {title} {\bibinfo {title} {Quantum theory of cavity-assisted sideband
  cooling of mechanical motion},\ }\href
  {https://doi.org/10.1103/PhysRevLett.99.093902} {\bibfield  {journal}
  {\bibinfo  {journal} {Phys. Rev. Lett.}\ }\textbf {\bibinfo {volume} {99}},\
  \bibinfo {pages} {093902} (\bibinfo {year} {2007})}\BibitemShut {NoStop}%
\bibitem [{\citenamefont {Liu}\ \emph {et~al.}(2013)\citenamefont {Liu},
  \citenamefont {Xiao}, \citenamefont {Luan},\ and\ \citenamefont
  {Wong}}]{PhysRevLett.110.153606}%
  \BibitemOpen
  \bibfield  {author} {\bibinfo {author} {\bibfnamefont {Y.-C.}\ \bibnamefont
  {Liu}}, \bibinfo {author} {\bibfnamefont {Y.-F.}\ \bibnamefont {Xiao}},
  \bibinfo {author} {\bibfnamefont {X.}~\bibnamefont {Luan}},\ and\ \bibinfo
  {author} {\bibfnamefont {C.~W.}\ \bibnamefont {Wong}},\ }\bibfield  {title}
  {\bibinfo {title} {Dynamic dissipative cooling of a mechanical resonator in
  strong coupling optomechanics},\ }\href
  {https://doi.org/10.1103/PhysRevLett.110.153606} {\bibfield  {journal}
  {\bibinfo  {journal} {Phys. Rev. Lett.}\ }\textbf {\bibinfo {volume} {110}},\
  \bibinfo {pages} {153606} (\bibinfo {year} {2013})}\BibitemShut {NoStop}%
\bibitem [{\citenamefont {Liu}\ \emph {et~al.}(2015)\citenamefont {Liu},
  \citenamefont {Xiao}, \citenamefont {Luan}, \citenamefont {Gong},\ and\
  \citenamefont {Wong}}]{PhysRevA.91.033818}%
  \BibitemOpen
  \bibfield  {author} {\bibinfo {author} {\bibfnamefont {Y.-C.}\ \bibnamefont
  {Liu}}, \bibinfo {author} {\bibfnamefont {Y.-F.}\ \bibnamefont {Xiao}},
  \bibinfo {author} {\bibfnamefont {X.}~\bibnamefont {Luan}}, \bibinfo {author}
  {\bibfnamefont {Q.}~\bibnamefont {Gong}},\ and\ \bibinfo {author}
  {\bibfnamefont {C.~W.}\ \bibnamefont {Wong}},\ }\bibfield  {title} {\bibinfo
  {title} {Coupled cavities for motional ground-state cooling and strong
  optomechanical coupling},\ }\href
  {https://doi.org/10.1103/PhysRevA.91.033818} {\bibfield  {journal} {\bibinfo
  {journal} {Phys. Rev. A}\ }\textbf {\bibinfo {volume} {91}},\ \bibinfo
  {pages} {033818} (\bibinfo {year} {2015})}\BibitemShut {NoStop}%
\bibitem [{\citenamefont {Park}\ and\ \citenamefont {Wang}(2009)}]{Park2009}%
  \BibitemOpen
  \bibfield  {author} {\bibinfo {author} {\bibfnamefont {Y.-S.}\ \bibnamefont
  {Park}}\ and\ \bibinfo {author} {\bibfnamefont {H.}~\bibnamefont {Wang}},\
  }\bibfield  {title} {\bibinfo {title} {Resolved-sideband and cryogenic
  cooling of an optomechanical resonator},\ }\href
  {https://doi.org/10.1038/nphys1303} {\bibfield  {journal} {\bibinfo
  {journal} {Nat. Phys.}\ }\textbf {\bibinfo {volume} {5}},\ \bibinfo {pages}
  {489} (\bibinfo {year} {2009})}\BibitemShut {NoStop}%
\bibitem [{\citenamefont {Rocheleau}\ \emph {et~al.}(2010)\citenamefont
  {Rocheleau}, \citenamefont {Ndukum}, \citenamefont {Macklin}, \citenamefont
  {Hertzberg}, \citenamefont {Clerk},\ and\ \citenamefont
  {Schwab}}]{Rocheleau2010}%
  \BibitemOpen
  \bibfield  {author} {\bibinfo {author} {\bibfnamefont {T.}~\bibnamefont
  {Rocheleau}}, \bibinfo {author} {\bibfnamefont {T.}~\bibnamefont {Ndukum}},
  \bibinfo {author} {\bibfnamefont {C.}~\bibnamefont {Macklin}}, \bibinfo
  {author} {\bibfnamefont {J.~B.}\ \bibnamefont {Hertzberg}}, \bibinfo {author}
  {\bibfnamefont {A.~A.}\ \bibnamefont {Clerk}},\ and\ \bibinfo {author}
  {\bibfnamefont {K.~C.}\ \bibnamefont {Schwab}},\ }\bibfield  {title}
  {\bibinfo {title} {Preparation and detection of a mechanical resonator near
  the ground state of motion},\ }\href {https://doi.org/10.1038/nature08681}
  {\bibfield  {journal} {\bibinfo  {journal} {Nature}\ }\textbf {\bibinfo
  {volume} {463}},\ \bibinfo {pages} {72} (\bibinfo {year} {2010})}\BibitemShut
  {NoStop}%
\bibitem [{\citenamefont {Rivi\`ere}\ \emph {et~al.}(2011)\citenamefont
  {Rivi\`ere}, \citenamefont {Del\'eglise}, \citenamefont {Weis}, \citenamefont
  {Gavartin}, \citenamefont {Arcizet}, \citenamefont {Schliesser},\ and\
  \citenamefont {Kippenberg}}]{PhysRevA.83.063835}%
  \BibitemOpen
  \bibfield  {author} {\bibinfo {author} {\bibfnamefont {R.}~\bibnamefont
  {Rivi\`ere}}, \bibinfo {author} {\bibfnamefont {S.}~\bibnamefont
  {Del\'eglise}}, \bibinfo {author} {\bibfnamefont {S.}~\bibnamefont {Weis}},
  \bibinfo {author} {\bibfnamefont {E.}~\bibnamefont {Gavartin}}, \bibinfo
  {author} {\bibfnamefont {O.}~\bibnamefont {Arcizet}}, \bibinfo {author}
  {\bibfnamefont {A.}~\bibnamefont {Schliesser}},\ and\ \bibinfo {author}
  {\bibfnamefont {T.~J.}\ \bibnamefont {Kippenberg}},\ }\bibfield  {title}
  {\bibinfo {title} {Optomechanical sideband cooling of a micromechanical
  oscillator close to the quantum ground state},\ }\href
  {https://doi.org/10.1103/PhysRevA.83.063835} {\bibfield  {journal} {\bibinfo
  {journal} {Phys. Rev. A}\ }\textbf {\bibinfo {volume} {83}},\ \bibinfo
  {pages} {063835} (\bibinfo {year} {2011})}\BibitemShut {NoStop}%
\bibitem [{\citenamefont {Teufel}\ \emph {et~al.}(2011)\citenamefont {Teufel},
  \citenamefont {Donner}, \citenamefont {Li}, \citenamefont {Harlow},
  \citenamefont {Allman}, \citenamefont {Cicak}, \citenamefont {Sirois},
  \citenamefont {Whittaker}, \citenamefont {Lehnert},\ and\ \citenamefont
  {Simmonds}}]{Teufel2011}%
  \BibitemOpen
  \bibfield  {author} {\bibinfo {author} {\bibfnamefont {J.~D.}\ \bibnamefont
  {Teufel}}, \bibinfo {author} {\bibfnamefont {T.}~\bibnamefont {Donner}},
  \bibinfo {author} {\bibfnamefont {D.}~\bibnamefont {Li}}, \bibinfo {author}
  {\bibfnamefont {J.~W.}\ \bibnamefont {Harlow}}, \bibinfo {author}
  {\bibfnamefont {M.~S.}\ \bibnamefont {Allman}}, \bibinfo {author}
  {\bibfnamefont {K.}~\bibnamefont {Cicak}}, \bibinfo {author} {\bibfnamefont
  {A.~J.}\ \bibnamefont {Sirois}}, \bibinfo {author} {\bibfnamefont {J.~D.}\
  \bibnamefont {Whittaker}}, \bibinfo {author} {\bibfnamefont {K.~W.}\
  \bibnamefont {Lehnert}},\ and\ \bibinfo {author} {\bibfnamefont {R.~W.}\
  \bibnamefont {Simmonds}},\ }\bibfield  {title} {\bibinfo {title} {Sideband
  cooling of micromechanical motion to the quantum ground state},\ }\href
  {https://doi.org/10.1038/nature10261} {\bibfield  {journal} {\bibinfo
  {journal} {Nature}\ }\textbf {\bibinfo {volume} {475}},\ \bibinfo {pages}
  {359} (\bibinfo {year} {2011})}\BibitemShut {NoStop}%
\bibitem [{\citenamefont {Chan}\ \emph {et~al.}(2011)\citenamefont {Chan},
  \citenamefont {Alegre}, \citenamefont {Safavi-Naeini}, \citenamefont {Hill},
  \citenamefont {Krause}, \citenamefont {Gr{\"o}blacher}, \citenamefont
  {Aspelmeyer},\ and\ \citenamefont {Painter}}]{Chan2011}%
  \BibitemOpen
  \bibfield  {author} {\bibinfo {author} {\bibfnamefont {J.}~\bibnamefont
  {Chan}}, \bibinfo {author} {\bibfnamefont {T.~P.~M.}\ \bibnamefont {Alegre}},
  \bibinfo {author} {\bibfnamefont {A.~H.}\ \bibnamefont {Safavi-Naeini}},
  \bibinfo {author} {\bibfnamefont {J.~T.}\ \bibnamefont {Hill}}, \bibinfo
  {author} {\bibfnamefont {A.}~\bibnamefont {Krause}}, \bibinfo {author}
  {\bibfnamefont {S.}~\bibnamefont {Gr{\"o}blacher}}, \bibinfo {author}
  {\bibfnamefont {M.}~\bibnamefont {Aspelmeyer}},\ and\ \bibinfo {author}
  {\bibfnamefont {O.}~\bibnamefont {Painter}},\ }\bibfield  {title} {\bibinfo
  {title} {Laser cooling of a nanomechanical oscillator into its quantum ground
  state},\ }\href {https://doi.org/10.1038/nature10461} {\bibfield  {journal}
  {\bibinfo  {journal} {Nature}\ }\textbf {\bibinfo {volume} {478}},\ \bibinfo
  {pages} {89} (\bibinfo {year} {2011})}\BibitemShut {NoStop}%
\bibitem [{\citenamefont {Verhagen}\ \emph {et~al.}(2012)\citenamefont
  {Verhagen}, \citenamefont {Del{\'e}glise}, \citenamefont {Weis},
  \citenamefont {Schliesser},\ and\ \citenamefont {Kippenberg}}]{Verhagen2012}%
  \BibitemOpen
  \bibfield  {author} {\bibinfo {author} {\bibfnamefont {E.}~\bibnamefont
  {Verhagen}}, \bibinfo {author} {\bibfnamefont {S.}~\bibnamefont
  {Del{\'e}glise}}, \bibinfo {author} {\bibfnamefont {S.}~\bibnamefont {Weis}},
  \bibinfo {author} {\bibfnamefont {A.}~\bibnamefont {Schliesser}},\ and\
  \bibinfo {author} {\bibfnamefont {T.~J.}\ \bibnamefont {Kippenberg}},\
  }\bibfield  {title} {\bibinfo {title} {Quantum-coherent coupling of a
  mechanical oscillator to an optical cavity mode},\ }\href
  {https://doi.org/10.1038/nature10787} {\bibfield  {journal} {\bibinfo
  {journal} {Nature}\ }\textbf {\bibinfo {volume} {482}},\ \bibinfo {pages}
  {63} (\bibinfo {year} {2012})}\BibitemShut {NoStop}%
\bibitem [{\citenamefont {Guo}\ \emph {et~al.}(2019)\citenamefont {Guo},
  \citenamefont {Norte},\ and\ \citenamefont
  {Gr\"oblacher}}]{PhysRevLett.123.223602}%
  \BibitemOpen
  \bibfield  {author} {\bibinfo {author} {\bibfnamefont {J.}~\bibnamefont
  {Guo}}, \bibinfo {author} {\bibfnamefont {R.}~\bibnamefont {Norte}},\ and\
  \bibinfo {author} {\bibfnamefont {S.}~\bibnamefont {Gr\"oblacher}},\
  }\bibfield  {title} {\bibinfo {title} {Feedback cooling of a room temperature
  mechanical oscillator close to its motional ground state},\ }\href
  {https://doi.org/10.1103/PhysRevLett.123.223602} {\bibfield  {journal}
  {\bibinfo  {journal} {Phys. Rev. Lett.}\ }\textbf {\bibinfo {volume} {123}},\
  \bibinfo {pages} {223602} (\bibinfo {year} {2019})}\BibitemShut {NoStop}%
\bibitem [{\citenamefont {Whittle}\ \emph {et~al.}(2021)\citenamefont
  {Whittle}, \citenamefont {Hall} \emph {et~al.}}]{Whittle1333}%
  \BibitemOpen
  \bibfield  {author} {\bibinfo {author} {\bibfnamefont {C.}~\bibnamefont
  {Whittle}}, \bibinfo {author} {\bibfnamefont {E.~D.}\ \bibnamefont {Hall}},
  \emph {et~al.},\ }\bibfield  {title} {\bibinfo {title} {Approaching the
  motional ground state of a 10-kg object},\ }\href
  {https://doi.org/10.1126/science.abh2634} {\bibfield  {journal} {\bibinfo
  {journal} {Science}\ }\textbf {\bibinfo {volume} {372}},\ \bibinfo {pages}
  {1333} (\bibinfo {year} {2021})}\BibitemShut {NoStop}%
\bibitem [{\citenamefont {Heinrich}\ \emph {et~al.}(2011)\citenamefont
  {Heinrich}, \citenamefont {Ludwig}, \citenamefont {Qian}, \citenamefont
  {Kubala},\ and\ \citenamefont {Marquardt}}]{PhysRevLett.107.043603}%
  \BibitemOpen
  \bibfield  {author} {\bibinfo {author} {\bibfnamefont {G.}~\bibnamefont
  {Heinrich}}, \bibinfo {author} {\bibfnamefont {M.}~\bibnamefont {Ludwig}},
  \bibinfo {author} {\bibfnamefont {J.}~\bibnamefont {Qian}}, \bibinfo {author}
  {\bibfnamefont {B.}~\bibnamefont {Kubala}},\ and\ \bibinfo {author}
  {\bibfnamefont {F.}~\bibnamefont {Marquardt}},\ }\bibfield  {title} {\bibinfo
  {title} {Collective dynamics in optomechanical arrays},\ }\href
  {https://doi.org/10.1103/PhysRevLett.107.043603} {\bibfield  {journal}
  {\bibinfo  {journal} {Phys. Rev. Lett.}\ }\textbf {\bibinfo {volume} {107}},\
  \bibinfo {pages} {043603} (\bibinfo {year} {2011})}\BibitemShut {NoStop}%
\bibitem [{\citenamefont {Ludwig}\ and\ \citenamefont
  {Marquardt}(2013)}]{PhysRevLett.111.073603}%
  \BibitemOpen
  \bibfield  {author} {\bibinfo {author} {\bibfnamefont {M.}~\bibnamefont
  {Ludwig}}\ and\ \bibinfo {author} {\bibfnamefont {F.}~\bibnamefont
  {Marquardt}},\ }\bibfield  {title} {\bibinfo {title} {Quantum many-body
  dynamics in optomechanical arrays},\ }\href
  {https://doi.org/10.1103/PhysRevLett.111.073603} {\bibfield  {journal}
  {\bibinfo  {journal} {Phys. Rev. Lett.}\ }\textbf {\bibinfo {volume} {111}},\
  \bibinfo {pages} {073603} (\bibinfo {year} {2013})}\BibitemShut {NoStop}%
\bibitem [{\citenamefont {Xuereb}\ \emph {et~al.}(2014)\citenamefont {Xuereb},
  \citenamefont {Genes}, \citenamefont {Pupillo}, \citenamefont {Paternostro},\
  and\ \citenamefont {Dantan}}]{PhysRevLett.112.133604}%
  \BibitemOpen
  \bibfield  {author} {\bibinfo {author} {\bibfnamefont {A.}~\bibnamefont
  {Xuereb}}, \bibinfo {author} {\bibfnamefont {C.}~\bibnamefont {Genes}},
  \bibinfo {author} {\bibfnamefont {G.}~\bibnamefont {Pupillo}}, \bibinfo
  {author} {\bibfnamefont {M.}~\bibnamefont {Paternostro}},\ and\ \bibinfo
  {author} {\bibfnamefont {A.}~\bibnamefont {Dantan}},\ }\bibfield  {title}
  {\bibinfo {title} {Reconfigurable long-range phonon dynamics in
  optomechanical arrays},\ }\href
  {https://doi.org/10.1103/PhysRevLett.112.133604} {\bibfield  {journal}
  {\bibinfo  {journal} {Phys. Rev. Lett.}\ }\textbf {\bibinfo {volume} {112}},\
  \bibinfo {pages} {133604} (\bibinfo {year} {2014})}\BibitemShut {NoStop}%
\bibitem [{\citenamefont {Stannigel}\ \emph {et~al.}(2012)\citenamefont
  {Stannigel}, \citenamefont {Komar}, \citenamefont {Habraken}, \citenamefont
  {Bennett}, \citenamefont {Lukin}, \citenamefont {Zoller},\ and\ \citenamefont
  {Rabl}}]{PhysRevLett.109.013603}%
  \BibitemOpen
  \bibfield  {author} {\bibinfo {author} {\bibfnamefont {K.}~\bibnamefont
  {Stannigel}}, \bibinfo {author} {\bibfnamefont {P.}~\bibnamefont {Komar}},
  \bibinfo {author} {\bibfnamefont {S.~J.~M.}\ \bibnamefont {Habraken}},
  \bibinfo {author} {\bibfnamefont {S.~D.}\ \bibnamefont {Bennett}}, \bibinfo
  {author} {\bibfnamefont {M.~D.}\ \bibnamefont {Lukin}}, \bibinfo {author}
  {\bibfnamefont {P.}~\bibnamefont {Zoller}},\ and\ \bibinfo {author}
  {\bibfnamefont {P.}~\bibnamefont {Rabl}},\ }\bibfield  {title} {\bibinfo
  {title} {Optomechanical quantum information processing with photons and
  phonons},\ }\href {https://doi.org/10.1103/PhysRevLett.109.013603} {\bibfield
   {journal} {\bibinfo  {journal} {Phys. Rev. Lett.}\ }\textbf {\bibinfo
  {volume} {109}},\ \bibinfo {pages} {013603} (\bibinfo {year}
  {2012})}\BibitemShut {NoStop}%
\bibitem [{\citenamefont {Fiore}\ \emph {et~al.}(2011)\citenamefont {Fiore},
  \citenamefont {Yang}, \citenamefont {Kuzyk}, \citenamefont {Barbour},
  \citenamefont {Tian},\ and\ \citenamefont {Wang}}]{PhysRevLett.107.133601}%
  \BibitemOpen
  \bibfield  {author} {\bibinfo {author} {\bibfnamefont {V.}~\bibnamefont
  {Fiore}}, \bibinfo {author} {\bibfnamefont {Y.}~\bibnamefont {Yang}},
  \bibinfo {author} {\bibfnamefont {M.~C.}\ \bibnamefont {Kuzyk}}, \bibinfo
  {author} {\bibfnamefont {R.}~\bibnamefont {Barbour}}, \bibinfo {author}
  {\bibfnamefont {L.}~\bibnamefont {Tian}},\ and\ \bibinfo {author}
  {\bibfnamefont {H.}~\bibnamefont {Wang}},\ }\bibfield  {title} {\bibinfo
  {title} {Storing optical information as a mechanical excitation in a silica
  optomechanical resonator},\ }\href
  {https://doi.org/10.1103/PhysRevLett.107.133601} {\bibfield  {journal}
  {\bibinfo  {journal} {Phys. Rev. Lett.}\ }\textbf {\bibinfo {volume} {107}},\
  \bibinfo {pages} {133601} (\bibinfo {year} {2011})}\BibitemShut {NoStop}%
\bibitem [{\citenamefont {Okamoto}\ \emph {et~al.}(2013)\citenamefont
  {Okamoto}, \citenamefont {Gourgout}, \citenamefont {Chang}, \citenamefont
  {Onomitsu}, \citenamefont {Mahboob}, \citenamefont {Chang},\ and\
  \citenamefont {Yamaguchi}}]{Okamoto2013}%
  \BibitemOpen
  \bibfield  {author} {\bibinfo {author} {\bibfnamefont {H.}~\bibnamefont
  {Okamoto}}, \bibinfo {author} {\bibfnamefont {A.}~\bibnamefont {Gourgout}},
  \bibinfo {author} {\bibfnamefont {C.-Y.}\ \bibnamefont {Chang}}, \bibinfo
  {author} {\bibfnamefont {K.}~\bibnamefont {Onomitsu}}, \bibinfo {author}
  {\bibfnamefont {I.}~\bibnamefont {Mahboob}}, \bibinfo {author} {\bibfnamefont
  {E.~Y.}\ \bibnamefont {Chang}},\ and\ \bibinfo {author} {\bibfnamefont
  {H.}~\bibnamefont {Yamaguchi}},\ }\bibfield  {title} {\bibinfo {title}
  {Coherent phonon manipulation in coupled mechanical resonators},\ }\href
  {https://doi.org/10.1038/nphys2665} {\bibfield  {journal} {\bibinfo
  {journal} {Nat. Phys.}\ }\textbf {\bibinfo {volume} {9}},\ \bibinfo {pages}
  {480} (\bibinfo {year} {2013})}\BibitemShut {NoStop}%
\bibitem [{\citenamefont {Wang}\ and\ \citenamefont
  {Clerk}(2012)}]{PhysRevLett.108.153603}%
  \BibitemOpen
  \bibfield  {author} {\bibinfo {author} {\bibfnamefont {Y.-D.}\ \bibnamefont
  {Wang}}\ and\ \bibinfo {author} {\bibfnamefont {A.~A.}\ \bibnamefont
  {Clerk}},\ }\bibfield  {title} {\bibinfo {title} {Using interference for high
  fidelity quantum state transfer in optomechanics},\ }\href
  {https://doi.org/10.1103/PhysRevLett.108.153603} {\bibfield  {journal}
  {\bibinfo  {journal} {Phys. Rev. Lett.}\ }\textbf {\bibinfo {volume} {108}},\
  \bibinfo {pages} {153603} (\bibinfo {year} {2012})}\BibitemShut {NoStop}%
\bibitem [{\citenamefont {Truitt}\ \emph {et~al.}(2007)\citenamefont {Truitt},
  \citenamefont {Hertzberg}, \citenamefont {Huang}, \citenamefont {Ekinci},\
  and\ \citenamefont {Schwab}}]{Truitt2007}%
  \BibitemOpen
  \bibfield  {author} {\bibinfo {author} {\bibfnamefont {P.~A.}\ \bibnamefont
  {Truitt}}, \bibinfo {author} {\bibfnamefont {J.~B.}\ \bibnamefont
  {Hertzberg}}, \bibinfo {author} {\bibfnamefont {C.~C.}\ \bibnamefont
  {Huang}}, \bibinfo {author} {\bibfnamefont {K.~L.}\ \bibnamefont {Ekinci}},\
  and\ \bibinfo {author} {\bibfnamefont {K.~C.}\ \bibnamefont {Schwab}},\
  }\bibfield  {title} {\bibinfo {title} {Efficient and sensitive capacitive
  readout of nanomechanical resonator arrays},\ }\href
  {https://doi.org/10.1021/nl062278g} {\bibfield  {journal} {\bibinfo
  {journal} {Nano Lett.}\ }\textbf {\bibinfo {volume} {7}},\ \bibinfo {pages}
  {120} (\bibinfo {year} {2007})}\BibitemShut {NoStop}%
\bibitem [{\citenamefont {Bargatin}\ \emph {et~al.}(2012)\citenamefont
  {Bargatin}, \citenamefont {Myers}, \citenamefont {Aldridge}, \citenamefont
  {Marcoux}, \citenamefont {Brianceau}, \citenamefont {Duraffourg},
  \citenamefont {Colinet}, \citenamefont {Hentz}, \citenamefont {Andreucci},\
  and\ \citenamefont {Roukes}}]{Bargatin2012}%
  \BibitemOpen
  \bibfield  {author} {\bibinfo {author} {\bibfnamefont {I.}~\bibnamefont
  {Bargatin}}, \bibinfo {author} {\bibfnamefont {E.~B.}\ \bibnamefont {Myers}},
  \bibinfo {author} {\bibfnamefont {J.~S.}\ \bibnamefont {Aldridge}}, \bibinfo
  {author} {\bibfnamefont {C.}~\bibnamefont {Marcoux}}, \bibinfo {author}
  {\bibfnamefont {P.}~\bibnamefont {Brianceau}}, \bibinfo {author}
  {\bibfnamefont {L.}~\bibnamefont {Duraffourg}}, \bibinfo {author}
  {\bibfnamefont {E.}~\bibnamefont {Colinet}}, \bibinfo {author} {\bibfnamefont
  {S.}~\bibnamefont {Hentz}}, \bibinfo {author} {\bibfnamefont
  {P.}~\bibnamefont {Andreucci}},\ and\ \bibinfo {author} {\bibfnamefont
  {M.~L.}\ \bibnamefont {Roukes}},\ }\bibfield  {title} {\bibinfo {title}
  {Large-scale integration of nanoelectromechanical systems for gas sensing
  applications},\ }\href {https://doi.org/10.1021/nl2037479} {\bibfield
  {journal} {\bibinfo  {journal} {Nano Lett.}\ }\textbf {\bibinfo {volume}
  {12}},\ \bibinfo {pages} {1269} (\bibinfo {year} {2012})}\BibitemShut
  {NoStop}%
\bibitem [{\citenamefont {Rabl}\ \emph {et~al.}(2010)\citenamefont {Rabl},
  \citenamefont {Kolkowitz}, \citenamefont {Koppens}, \citenamefont {Harris},
  \citenamefont {Zoller},\ and\ \citenamefont {Lukin}}]{Rabl2010}%
  \BibitemOpen
  \bibfield  {author} {\bibinfo {author} {\bibfnamefont {P.}~\bibnamefont
  {Rabl}}, \bibinfo {author} {\bibfnamefont {S.~J.}\ \bibnamefont {Kolkowitz}},
  \bibinfo {author} {\bibfnamefont {F.~H.~L.}\ \bibnamefont {Koppens}},
  \bibinfo {author} {\bibfnamefont {J.~G.~E.}\ \bibnamefont {Harris}}, \bibinfo
  {author} {\bibfnamefont {P.}~\bibnamefont {Zoller}},\ and\ \bibinfo {author}
  {\bibfnamefont {M.~D.}\ \bibnamefont {Lukin}},\ }\bibfield  {title} {\bibinfo
  {title} {A quantum spin transducer based on nanoelectromechanical resonator
  arrays},\ }\href {https://doi.org/10.1038/nphys1679} {\bibfield  {journal}
  {\bibinfo  {journal} {Nat. Phys.}\ }\textbf {\bibinfo {volume} {6}},\
  \bibinfo {pages} {602} (\bibinfo {year} {2010})}\BibitemShut {NoStop}%
\bibitem [{\citenamefont {Huang}\ \emph {et~al.}(2013)\citenamefont {Huang},
  \citenamefont {Wang}, \citenamefont {Zhou}, \citenamefont {Wang},
  \citenamefont {Ju}, \citenamefont {Wang}, \citenamefont {Shen}, \citenamefont
  {Duan},\ and\ \citenamefont {Du}}]{PhysRevLett.110.227202}%
  \BibitemOpen
  \bibfield  {author} {\bibinfo {author} {\bibfnamefont {P.}~\bibnamefont
  {Huang}}, \bibinfo {author} {\bibfnamefont {P.}~\bibnamefont {Wang}},
  \bibinfo {author} {\bibfnamefont {J.}~\bibnamefont {Zhou}}, \bibinfo {author}
  {\bibfnamefont {Z.}~\bibnamefont {Wang}}, \bibinfo {author} {\bibfnamefont
  {C.}~\bibnamefont {Ju}}, \bibinfo {author} {\bibfnamefont {Z.}~\bibnamefont
  {Wang}}, \bibinfo {author} {\bibfnamefont {Y.}~\bibnamefont {Shen}}, \bibinfo
  {author} {\bibfnamefont {C.}~\bibnamefont {Duan}},\ and\ \bibinfo {author}
  {\bibfnamefont {J.}~\bibnamefont {Du}},\ }\bibfield  {title} {\bibinfo
  {title} {Demonstration of motion transduction based on parametrically coupled
  mechanical resonators},\ }\href
  {https://doi.org/10.1103/PhysRevLett.110.227202} {\bibfield  {journal}
  {\bibinfo  {journal} {Phys. Rev. Lett.}\ }\textbf {\bibinfo {volume} {110}},\
  \bibinfo {pages} {227202} (\bibinfo {year} {2013})}\BibitemShut {NoStop}%
\bibitem [{\citenamefont {Huang}\ \emph {et~al.}(2016)\citenamefont {Huang},
  \citenamefont {Zhang}, \citenamefont {Zhou}, \citenamefont {Tian},
  \citenamefont {Yin}, \citenamefont {Duan},\ and\ \citenamefont
  {Du}}]{PhysRevLett.117.017701}%
  \BibitemOpen
  \bibfield  {author} {\bibinfo {author} {\bibfnamefont {P.}~\bibnamefont
  {Huang}}, \bibinfo {author} {\bibfnamefont {L.}~\bibnamefont {Zhang}},
  \bibinfo {author} {\bibfnamefont {J.}~\bibnamefont {Zhou}}, \bibinfo {author}
  {\bibfnamefont {T.}~\bibnamefont {Tian}}, \bibinfo {author} {\bibfnamefont
  {P.}~\bibnamefont {Yin}}, \bibinfo {author} {\bibfnamefont {C.}~\bibnamefont
  {Duan}},\ and\ \bibinfo {author} {\bibfnamefont {J.}~\bibnamefont {Du}},\
  }\bibfield  {title} {\bibinfo {title} {Nonreciprocal radio frequency
  transduction in a parametric mechanical artificial lattice},\ }\href
  {https://doi.org/10.1103/PhysRevLett.117.017701} {\bibfield  {journal}
  {\bibinfo  {journal} {Phys. Rev. Lett.}\ }\textbf {\bibinfo {volume} {117}},\
  \bibinfo {pages} {017701} (\bibinfo {year} {2016})}\BibitemShut {NoStop}%
\bibitem [{\citenamefont {Genes}\ \emph {et~al.}(2008)\citenamefont {Genes},
  \citenamefont {Vitali},\ and\ \citenamefont {Tombesi}}]{Genes2008}%
  \BibitemOpen
  \bibfield  {author} {\bibinfo {author} {\bibfnamefont {C.}~\bibnamefont
  {Genes}}, \bibinfo {author} {\bibfnamefont {D.}~\bibnamefont {Vitali}},\ and\
  \bibinfo {author} {\bibfnamefont {P.}~\bibnamefont {Tombesi}},\ }\bibfield
  {title} {\bibinfo {title} {{Simultaneous cooling and entanglement of
  mechanical modes of a micromirror in an optical cavity}},\ }\href
  {https://doi.org/10.1088/1367-2630/10/9/095009} {\bibfield  {journal}
  {\bibinfo  {journal} {New J. Phys.}\ }\textbf {\bibinfo {volume} {10}},\
  \bibinfo {pages} {095009} (\bibinfo {year} {2008})}\BibitemShut {NoStop}%
\bibitem [{\citenamefont {Sommer}\ and\ \citenamefont
  {Genes}(2019)}]{Sommer2019}%
  \BibitemOpen
  \bibfield  {author} {\bibinfo {author} {\bibfnamefont {C.}~\bibnamefont
  {Sommer}}\ and\ \bibinfo {author} {\bibfnamefont {C.}~\bibnamefont {Genes}},\
  }\bibfield  {title} {\bibinfo {title} {{Partial Optomechanical Refrigeration
  via Multimode Cold-Damping Feedback}},\ }\href
  {https://doi.org/10.1103/PhysRevLett.123.203605} {\bibfield  {journal}
  {\bibinfo  {journal} {Phys. Rev. Lett.}\ }\textbf {\bibinfo {volume} {123}},\
  \bibinfo {pages} {203605} (\bibinfo {year} {2019})}\BibitemShut {NoStop}%
\bibitem [{\citenamefont {Shkarin}\ \emph {et~al.}(2014)\citenamefont
  {Shkarin}, \citenamefont {Flowers-Jacobs}, \citenamefont {Hoch},
  \citenamefont {Kashkanova}, \citenamefont {Deutsch}, \citenamefont
  {Reichel},\ and\ \citenamefont {Harris}}]{Shkarin2014}%
  \BibitemOpen
  \bibfield  {author} {\bibinfo {author} {\bibfnamefont {A.~B.}\ \bibnamefont
  {Shkarin}}, \bibinfo {author} {\bibfnamefont {N.~E.}\ \bibnamefont
  {Flowers-Jacobs}}, \bibinfo {author} {\bibfnamefont {S.~W.}\ \bibnamefont
  {Hoch}}, \bibinfo {author} {\bibfnamefont {A.~D.}\ \bibnamefont
  {Kashkanova}}, \bibinfo {author} {\bibfnamefont {C.}~\bibnamefont {Deutsch}},
  \bibinfo {author} {\bibfnamefont {J.}~\bibnamefont {Reichel}},\ and\ \bibinfo
  {author} {\bibfnamefont {J.~G.~E.}\ \bibnamefont {Harris}},\ }\bibfield
  {title} {\bibinfo {title} {{Optically Mediated Hybridization between Two
  Mechanical Modes}},\ }\href {https://doi.org/10.1103/PhysRevLett.112.013602}
  {\bibfield  {journal} {\bibinfo  {journal} {Phys. Rev. Lett.}\ }\textbf
  {\bibinfo {volume} {112}},\ \bibinfo {pages} {013602} (\bibinfo {year}
  {2014})}\BibitemShut {NoStop}%
\bibitem [{\citenamefont {Ockeloen-Korppi}\ \emph {et~al.}(2019)\citenamefont
  {Ockeloen-Korppi}, \citenamefont {Gely}, \citenamefont {Damsk{\"{a}}gg},
  \citenamefont {Jenkins}, \citenamefont {Steele},\ and\ \citenamefont
  {Sillanp{\"{a}}{\"{a}}}}]{Ockeloen-Korppi2019}%
  \BibitemOpen
  \bibfield  {author} {\bibinfo {author} {\bibfnamefont {C.~F.}\ \bibnamefont
  {Ockeloen-Korppi}}, \bibinfo {author} {\bibfnamefont {M.~F.}\ \bibnamefont
  {Gely}}, \bibinfo {author} {\bibfnamefont {E.}~\bibnamefont
  {Damsk{\"{a}}gg}}, \bibinfo {author} {\bibfnamefont {M.}~\bibnamefont
  {Jenkins}}, \bibinfo {author} {\bibfnamefont {G.~A.}\ \bibnamefont
  {Steele}},\ and\ \bibinfo {author} {\bibfnamefont {M.~A.}\ \bibnamefont
  {Sillanp{\"{a}}{\"{a}}}},\ }\bibfield  {title} {\bibinfo {title} {{Sideband
  cooling of nearly degenerate micromechanical oscillators in a multimode
  optomechanical system}},\ }\href {https://doi.org/10.1103/PhysRevA.99.023826}
  {\bibfield  {journal} {\bibinfo  {journal} {Phys. Rev. A}\ }\textbf {\bibinfo
  {volume} {99}},\ \bibinfo {pages} {023826} (\bibinfo {year}
  {2019})}\BibitemShut {NoStop}%
\bibitem [{\citenamefont {Lai}\ \emph {et~al.}(2018)\citenamefont {Lai},
  \citenamefont {Zou}, \citenamefont {Hou}, \citenamefont {Xiao},\ and\
  \citenamefont {Liao}}]{Lai2018}%
  \BibitemOpen
  \bibfield  {author} {\bibinfo {author} {\bibfnamefont {D.-G.}\ \bibnamefont
  {Lai}}, \bibinfo {author} {\bibfnamefont {F.}~\bibnamefont {Zou}}, \bibinfo
  {author} {\bibfnamefont {B.-P.}\ \bibnamefont {Hou}}, \bibinfo {author}
  {\bibfnamefont {Y.-F.}\ \bibnamefont {Xiao}},\ and\ \bibinfo {author}
  {\bibfnamefont {J.-Q.}\ \bibnamefont {Liao}},\ }\bibfield  {title} {\bibinfo
  {title} {{Simultaneous cooling of coupled mechanical resonators in cavity
  optomechanics}},\ }\href {https://doi.org/10.1103/PhysRevA.98.023860}
  {\bibfield  {journal} {\bibinfo  {journal} {Phys. Rev. A}\ }\textbf {\bibinfo
  {volume} {98}},\ \bibinfo {pages} {023860} (\bibinfo {year}
  {2018})}\BibitemShut {NoStop}%
\bibitem [{\citenamefont {Zhang}\ \emph {et~al.}(2019)\citenamefont {Zhang},
  \citenamefont {Zhou}, \citenamefont {Guo},\ and\ \citenamefont
  {Yi}}]{Zhang2019_dissipative}%
  \BibitemOpen
  \bibfield  {author} {\bibinfo {author} {\bibfnamefont {X.~Y.}\ \bibnamefont
  {Zhang}}, \bibinfo {author} {\bibfnamefont {Y.~H.}\ \bibnamefont {Zhou}},
  \bibinfo {author} {\bibfnamefont {Y.~Q.}\ \bibnamefont {Guo}},\ and\ \bibinfo
  {author} {\bibfnamefont {X.~X.}\ \bibnamefont {Yi}},\ }\bibfield  {title}
  {\bibinfo {title} {{Simultaneous cooling of two mechanical oscillators in
  dissipatively coupled optomechanical systems}},\ }\href
  {https://doi.org/10.1103/PhysRevA.100.023807} {\bibfield  {journal} {\bibinfo
   {journal} {Phys. Rev. A}\ }\textbf {\bibinfo {volume} {100}},\ \bibinfo
  {pages} {023807} (\bibinfo {year} {2019})}\BibitemShut {NoStop}%
\bibitem [{\citenamefont {Lai}\ \emph {et~al.}(2020)\citenamefont {Lai},
  \citenamefont {Huang}, \citenamefont {Yin}, \citenamefont {Hou},
  \citenamefont {Li}, \citenamefont {Vitali}, \citenamefont {Nori},\ and\
  \citenamefont {Liao}}]{Lai2020_cool}%
  \BibitemOpen
  \bibfield  {author} {\bibinfo {author} {\bibfnamefont {D.-G.}\ \bibnamefont
  {Lai}}, \bibinfo {author} {\bibfnamefont {J.-F.}\ \bibnamefont {Huang}},
  \bibinfo {author} {\bibfnamefont {X.-L.}\ \bibnamefont {Yin}}, \bibinfo
  {author} {\bibfnamefont {B.-P.}\ \bibnamefont {Hou}}, \bibinfo {author}
  {\bibfnamefont {W.}~\bibnamefont {Li}}, \bibinfo {author} {\bibfnamefont
  {D.}~\bibnamefont {Vitali}}, \bibinfo {author} {\bibfnamefont
  {F.}~\bibnamefont {Nori}},\ and\ \bibinfo {author} {\bibfnamefont {J.-Q.}\
  \bibnamefont {Liao}},\ }\bibfield  {title} {\bibinfo {title} {{Nonreciprocal
  ground-state cooling of multiple mechanical resonators}},\ }\href
  {https://doi.org/10.1103/PhysRevA.102.011502} {\bibfield  {journal} {\bibinfo
   {journal} {Phys. Rev. A}\ }\textbf {\bibinfo {volume} {102}},\ \bibinfo
  {pages} {011502} (\bibinfo {year} {2020})}\BibitemShut {NoStop}%
\bibitem [{\citenamefont {Habraken}\ \emph {et~al.}(2012)\citenamefont
  {Habraken}, \citenamefont {Stannigel}, \citenamefont {Lukin}, \citenamefont
  {Zoller},\ and\ \citenamefont {Rabl}}]{Habraken_2012}%
  \BibitemOpen
  \bibfield  {author} {\bibinfo {author} {\bibfnamefont {S.~J.~M.}\
  \bibnamefont {Habraken}}, \bibinfo {author} {\bibfnamefont {K.}~\bibnamefont
  {Stannigel}}, \bibinfo {author} {\bibfnamefont {M.~D.}\ \bibnamefont
  {Lukin}}, \bibinfo {author} {\bibfnamefont {P.}~\bibnamefont {Zoller}},\ and\
  \bibinfo {author} {\bibfnamefont {P.}~\bibnamefont {Rabl}},\ }\bibfield
  {title} {\bibinfo {title} {Continuous mode cooling and phonon routers for
  phononic quantum networks},\ }\href
  {https://doi.org/10.1088/1367-2630/14/11/115004} {\bibfield  {journal}
  {\bibinfo  {journal} {New J. Phys.}\ }\textbf {\bibinfo {volume} {14}},\
  \bibinfo {pages} {115004} (\bibinfo {year} {2012})}\BibitemShut {NoStop}%
\bibitem [{\citenamefont {Kim}\ \emph {et~al.}(2017)\citenamefont {Kim},
  \citenamefont {Xu}, \citenamefont {Taylor},\ and\ \citenamefont
  {Bahl}}]{Kim2017}%
  \BibitemOpen
  \bibfield  {author} {\bibinfo {author} {\bibfnamefont {S.}~\bibnamefont
  {Kim}}, \bibinfo {author} {\bibfnamefont {X.}~\bibnamefont {Xu}}, \bibinfo
  {author} {\bibfnamefont {J.~M.}\ \bibnamefont {Taylor}},\ and\ \bibinfo
  {author} {\bibfnamefont {G.}~\bibnamefont {Bahl}},\ }\bibfield  {title}
  {\bibinfo {title} {Dynamically induced robust phonon transport and chiral
  cooling in an optomechanical system},\ }\href
  {https://doi.org/10.1038/s41467-017-00247-7} {\bibfield  {journal} {\bibinfo
  {journal} {Nat. Commun.}\ }\textbf {\bibinfo {volume} {8}},\ \bibinfo {pages}
  {205} (\bibinfo {year} {2017})}\BibitemShut {NoStop}%
\bibitem [{\citenamefont {Xu}\ \emph {et~al.}(2019)\citenamefont {Xu},
  \citenamefont {Jiang}, \citenamefont {Clerk},\ and\ \citenamefont
  {Harris}}]{Xu2019}%
  \BibitemOpen
  \bibfield  {author} {\bibinfo {author} {\bibfnamefont {H.}~\bibnamefont
  {Xu}}, \bibinfo {author} {\bibfnamefont {L.}~\bibnamefont {Jiang}}, \bibinfo
  {author} {\bibfnamefont {A.~A.}\ \bibnamefont {Clerk}},\ and\ \bibinfo
  {author} {\bibfnamefont {J.~G.}\ \bibnamefont {Harris}},\ }\bibfield  {title}
  {\bibinfo {title} {{Nonreciprocal control and cooling of phonon modes in an
  optomechanical system}},\ }\href {https://doi.org/10.1038/s41586-019-1061-2}
  {\bibfield  {journal} {\bibinfo  {journal} {Nature}\ }\textbf {\bibinfo
  {volume} {568}},\ \bibinfo {pages} {65} (\bibinfo {year} {2019})}\BibitemShut
  {NoStop}%
\bibitem [{\citenamefont {Boller}\ \emph {et~al.}(1991)\citenamefont {Boller},
  \citenamefont {Imamo\ifmmode~\breve{g}\else \u{g}\fi{}lu},\ and\
  \citenamefont {Harris}}]{PhysRevLett.66.2593}%
  \BibitemOpen
  \bibfield  {author} {\bibinfo {author} {\bibfnamefont {K.-J.}\ \bibnamefont
  {Boller}}, \bibinfo {author} {\bibfnamefont {A.}~\bibnamefont
  {Imamo\ifmmode~\breve{g}\else \u{g}\fi{}lu}},\ and\ \bibinfo {author}
  {\bibfnamefont {S.~E.}\ \bibnamefont {Harris}},\ }\bibfield  {title}
  {\bibinfo {title} {Observation of electromagnetically induced transparency},\
  }\href {https://doi.org/10.1103/PhysRevLett.66.2593} {\bibfield  {journal}
  {\bibinfo  {journal} {Phys. Rev. Lett.}\ }\textbf {\bibinfo {volume} {66}},\
  \bibinfo {pages} {2593} (\bibinfo {year} {1991})}\BibitemShut {NoStop}%
\bibitem [{\citenamefont {Purdy}\ \emph {et~al.}(2012)\citenamefont {Purdy},
  \citenamefont {Peterson}, \citenamefont {Yu},\ and\ \citenamefont
  {Regal}}]{Purdy_2012}%
  \BibitemOpen
  \bibfield  {author} {\bibinfo {author} {\bibfnamefont {T.~P.}\ \bibnamefont
  {Purdy}}, \bibinfo {author} {\bibfnamefont {R.~W.}\ \bibnamefont {Peterson}},
  \bibinfo {author} {\bibfnamefont {P.-L.}\ \bibnamefont {Yu}},\ and\ \bibinfo
  {author} {\bibfnamefont {C.~A.}\ \bibnamefont {Regal}},\ }\bibfield  {title}
  {\bibinfo {title} {Cavity optomechanics with $\text{Si}_3\text{N}_4$
  membranes at cryogenic temperatures},\ }\href
  {https://doi.org/10.1088/1367-2630/14/11/115021} {\bibfield  {journal}
  {\bibinfo  {journal} {New J. Phys.}\ }\textbf {\bibinfo {volume} {14}},\
  \bibinfo {pages} {115021} (\bibinfo {year} {2012})}\BibitemShut {NoStop}%
\bibitem [{\citenamefont {Jayich}\ \emph {et~al.}(2008)\citenamefont {Jayich},
  \citenamefont {Sankey}, \citenamefont {Zwickl}, \citenamefont {Yang},
  \citenamefont {Thompson}, \citenamefont {Girvin}, \citenamefont {Clerk},
  \citenamefont {Marquardt},\ and\ \citenamefont {Harris}}]{Jayich_2008}%
  \BibitemOpen
  \bibfield  {author} {\bibinfo {author} {\bibfnamefont {A.~M.}\ \bibnamefont
  {Jayich}}, \bibinfo {author} {\bibfnamefont {J.~C.}\ \bibnamefont {Sankey}},
  \bibinfo {author} {\bibfnamefont {B.~M.}\ \bibnamefont {Zwickl}}, \bibinfo
  {author} {\bibfnamefont {C.}~\bibnamefont {Yang}}, \bibinfo {author}
  {\bibfnamefont {J.~D.}\ \bibnamefont {Thompson}}, \bibinfo {author}
  {\bibfnamefont {S.~M.}\ \bibnamefont {Girvin}}, \bibinfo {author}
  {\bibfnamefont {A.~A.}\ \bibnamefont {Clerk}}, \bibinfo {author}
  {\bibfnamefont {F.}~\bibnamefont {Marquardt}},\ and\ \bibinfo {author}
  {\bibfnamefont {J.~G.~E.}\ \bibnamefont {Harris}},\ }\bibfield  {title}
  {\bibinfo {title} {Dispersive optomechanics: a membrane inside a cavity},\
  }\href {https://doi.org/10.1088/1367-2630/10/9/095008} {\bibfield  {journal}
  {\bibinfo  {journal} {New Journal of Physics}\ }\textbf {\bibinfo {volume}
  {10}},\ \bibinfo {pages} {095008} (\bibinfo {year} {2008})}\BibitemShut
  {NoStop}%
\bibitem [{\citenamefont {Yu}\ \emph {et~al.}(2012)\citenamefont {Yu},
  \citenamefont {Purdy},\ and\ \citenamefont {Regal}}]{PhysRevLett.108.083603}%
  \BibitemOpen
  \bibfield  {author} {\bibinfo {author} {\bibfnamefont {P.-L.}\ \bibnamefont
  {Yu}}, \bibinfo {author} {\bibfnamefont {T.~P.}\ \bibnamefont {Purdy}},\ and\
  \bibinfo {author} {\bibfnamefont {C.~A.}\ \bibnamefont {Regal}},\ }\bibfield
  {title} {\bibinfo {title} {Control of material damping in high-$q$ membrane
  microresonators},\ }\href {https://doi.org/10.1103/PhysRevLett.108.083603}
  {\bibfield  {journal} {\bibinfo  {journal} {Phys. Rev. Lett.}\ }\textbf
  {\bibinfo {volume} {108}},\ \bibinfo {pages} {083603} (\bibinfo {year}
  {2012})}\BibitemShut {NoStop}%
\bibitem [{\citenamefont {Norte}\ \emph {et~al.}(2016)\citenamefont {Norte},
  \citenamefont {Moura},\ and\ \citenamefont
  {Gr\"oblacher}}]{PhysRevLett.116.147202}%
  \BibitemOpen
  \bibfield  {author} {\bibinfo {author} {\bibfnamefont {R.~A.}\ \bibnamefont
  {Norte}}, \bibinfo {author} {\bibfnamefont {J.~P.}\ \bibnamefont {Moura}},\
  and\ \bibinfo {author} {\bibfnamefont {S.}~\bibnamefont {Gr\"oblacher}},\
  }\bibfield  {title} {\bibinfo {title} {Mechanical resonators for quantum
  optomechanics experiments at room temperature},\ }\href
  {https://doi.org/10.1103/PhysRevLett.116.147202} {\bibfield  {journal}
  {\bibinfo  {journal} {Phys. Rev. Lett.}\ }\textbf {\bibinfo {volume} {116}},\
  \bibinfo {pages} {147202} (\bibinfo {year} {2016})}\BibitemShut {NoStop}%
\bibitem [{\citenamefont {Law}(1995)}]{PhysRevA.51.2537}%
  \BibitemOpen
  \bibfield  {author} {\bibinfo {author} {\bibfnamefont {C.~K.}\ \bibnamefont
  {Law}},\ }\bibfield  {title} {\bibinfo {title} {Interaction between a moving
  mirror and radiation pressure: A hamiltonian formulation},\ }\href
  {https://doi.org/10.1103/PhysRevA.51.2537} {\bibfield  {journal} {\bibinfo
  {journal} {Phys. Rev. A}\ }\textbf {\bibinfo {volume} {51}},\ \bibinfo
  {pages} {2537} (\bibinfo {year} {1995})}\BibitemShut {NoStop}%
\bibitem [{\citenamefont {Praxmeyer}\ and\ \citenamefont
  {Zloshchastiev}(2019)}]{Praxmeyer_2019}%
  \BibitemOpen
  \bibfield  {author} {\bibinfo {author} {\bibfnamefont {L.}~\bibnamefont
  {Praxmeyer}}\ and\ \bibinfo {author} {\bibfnamefont {K.~G.}\ \bibnamefont
  {Zloshchastiev}},\ }\bibfield  {title} {\bibinfo {title} {Master equation
  approach for non-hermitian quadratic hamiltonians: Original and phase space
  formulations},\ }\href {https://doi.org/10.1088/1742-6596/1194/1/012090}
  {\bibfield  {journal} {\bibinfo  {journal} {J. Phys. Conf. Ser.}\ }\textbf
  {\bibinfo {volume} {1194}},\ \bibinfo {pages} {012090} (\bibinfo {year}
  {2019})}\BibitemShut {NoStop}%
\end{thebibliography}%

\end{document}